\documentclass[twocolumn,traditabstract]{aa} 
\usepackage{txfonts}
\usepackage{color}
\usepackage{natbib}
\usepackage{multirow}
\usepackage{graphicx}
\usepackage{lscape}
\newcommand{\myrule}{\rule[-0.1cm]{0.cm}{0.5cm}} 
\begin{document} 

\title{A homogeneous analysis of disks around brown dwarfs
      \thanks{{\it Herschel} is an ESA space observatory with science instruments provided by European-led 
              Principal Investigator consortia and with important participation from NASA.}}

  \author{Y. Liu \inst{1, 2}           
          \and V. Joergens \inst{3, 4}
          \and A. Bayo \inst{5, 3}
          \and M. Nielbock \inst{3}
          \and H. Wang \inst{1, 2}
          }

  \institute{
             Purple Mountain Observatory, Chinese Academy of Sciences, Nanjing 210008, China
             \and
             Key Laboratory for Radio Astronomy, Chinese Academy of Sciences, 2 West Beijing Road, Nanjing 210008, China             
             \and
	     Max-Planck Institut f\"ur Astronomie, K\"onigstuhl~17, 69117 Heidelberg, Germany
	     \and  
             Universit\"at Heidelberg, Zentrum f\"ur Astronomie, Inst. f\"ur Theor. Astrophysik, Albert-Ueberle-Str. 2, 69120 Heidelberg, Germany
             \and
             Departamento de F\'isica y Astronom\'ia, Facultad de Ciencias, Universidad de Valpara\'iso, Av. Gran Breta\~na 1111, 5030 Casilla, Valpara\'iso, Chile
             }

   \authorrunning{Liu et al.}
   \titlerunning{Homogeneous analysis of disks around brown dwarfs}

   \abstract{We re-analyzed the {\it Herschel}/PACS data of a sample of 55 brown dwarfs (BDs) and very low mass stars with 
             spectral types ranging from M5.5 to L0. We investigated the dependence of disk structure on the mass 
             of the central object in the substellar regime based on a homogeneous analysis of {\it Herschel} data from  
             flux density measurements to spectral energy distribution (SED) modeling. The broadband SEDs were compiled by adding previous 
             photometry at shorter wavelengths and (sub-)millimeter data. We performed detailed SED analysis for the 46 
             targets that show infrared (IR) excess emission using radiative transfer models and evaluated the 
             constraints on the disk parameters through Bayesian inference. A systematic comparison between the derived disk 
             properties and those of sun-like stars shows that the disk flaring of BDs and very low mass stars is generally smaller than 
             that of their higher mass counterparts, the disk mass is orders of magnitude lower than the typical value found in 
             T Tauri stars, and the disk scale heights are comparable in both sun-like stars and BDs. We further divided our sample 
             into an early-type brown dwarf (ETBD) group and a late-type brown dwarf (LTBD) group by using spectral type 
             ($=\rm{M8}$) as the border criterion. We systematically compared the modeling results from Bayesian analysis between 
             these two groups, and found the trends of flaring index as a function of spectral type also present 
             in the substellar regime. The spectral type independence of the scale height is also seen between high-mass and very low-mass BDs.
             However, both the ETBD and LTBD groups feature a similar median disk mass of $1\times10^{-5}\,M_{\odot}$ and no clear trend is 
             visible in the distribution, probably due to the uncertainty in translating the far-IR photometry into disk mass, the 
             detection bias and the age difference among the sample. Unlike previous studies, our analysis is completely homogeneous 
             in {\it Herschel}/PACS data reduction and modeling with a statistically significant sample. Therefore, we present 
             evidence of stellar-mass-dependent disk structure down to the substellar mass regime, which is important for planet 
             formation models.}


   \maketitle

\section{Introduction}
How brown dwarfs (BDs) form is one of the main open questions in the field of star formation and remains 
a subject of debate although several scenarios have been proposed, for instance, a scaled down version of star 
formation processes, gravitational instabilities in disks and ejection of the stellar embryo \citep[e.g.,][]{reipurth2001, bate2003, stamatellos2011, chabrier2014}.
Similar to their higher mass counterparts, such as T Tauri stars, young BDs are shown to have circumstellar 
disks, producing substantial excess emission at wavelengths ranging from infrared (IR) to 
(sub)millimeter \citep[e.g.,][]{liun2003, scholz2006, bayo2012, harvey2012a, harvey2014, ricci2014, liu2015}. 
Disks are also found around very faint objects with masses down to the planetary regime, 
for example ${\rm Cha}\,110913{-}773444$ \citep[${\sim}8\,M_{\rm{Jup}}$,][]{luhman2005a},
LOri\,156 \citep[${\sim}23\,M_{\rm{Jup}}$,][]{bayo2012}, and OTS\,44 \citep[${\sim}12\,M_{\rm{Jup}}$,][]{luhman2005b,joergens2013}.
The phenomena of mass accretion and outflow, which are common by-products of the star formation process \citep[e.g.,][for reviews]{audard2014,frank2014},
have also been detected in young BDs \citep[e.g,][]{mohanty2005, phan2008, rigliaco2012, joergens2012}
even down to the planetary mass regime \citep[e.g.,][]{bayo2012, joergens2013}. The dust evolution in BD disks 
appears to follow a similar manner (i.e., grain growth, settling and crystallization) to that in T Tauri disks, although 
observations suggest different timescales of dust processing in disks around different stellar mass 
hosts \citep[e.g.,][]{apai2005,bouy2008,pascucci2009,riaz2012b}. These observations show that BDs resemble 
hydrogen-burning stars in many aspects during their early evolution, implying that they may form through the 
canonical star formation processes. 

Characterizing the physical properties of disks plays a crucial role in understanding the 
formation mechanisms of BDs \citep[e.g.,][]{reipurth2001,bate2003} and planets.
In addition, thorough comparisons between disk properties like the mass, flaring and scale height in 
different stellar mass regimes are very useful and important. 
So far, many works have been done in this direction.
Early disk studies in substellar regime are mainly based on {\it Spitzer} data.
For instance, \citet{szucs2010} investigated the IRAC and MIPS\,$24\,\mu{\rm{m}}$ photometry of ${\sim}200$ stars in the 
Chamaeleon I star formation region and found that disks around cooler objects are generally 
flatter than the case of earlier type stars. The {\it Herschel} space telescope has unprecedented sensitivity and 
angular resolution in the far-IR, enabling the detection of many faint disks at this wavelength domain for 
the first time. The {\it Herschel} data on one hand improve the determination of disk properties, 
and on the other hand provide an alternative way to estimate the disk mass of BDs.
\citet{harvey2012a} observed a large sample of low-mass stars and BDs with {\it Herschel}/PACS and found that the disk masses 
of low-mass stars and BDs extend to well below typical values found in T Tauri stars. The lower disk mass around cooler 
stars as revealed by far-IR measurements is consistent with the strong correlation between the stellar and disk masses given 
by \citet{andrewsr2013} from a millimeter survey of ${\rm Class\,\,II}$ sources in the Taurus molecular cloud. 
Several recent works obtained similar results, although the sample size and assumptions used to analyze the {\it Herschel} data 
differ \citep[e.g.,][]{alves2013,spezzi2013,olofsson2013,liu2015}. Note that the above disk comparisons were mostly
conducted between two stellar mass bins, i.e., the sun-like stars and the BDs including very low mass stars. These works
have yielded some evidence of spectral-type-dependent disk structure in the low stellar mass regime.

To date, however, there are seldom studies to investigate the dependency of disk properties on the spectral type (or the mass) 
of the central object within the substellar regime. It is quite unclear that whether the observed trend of disk properties 
as a function of spectral type is also valid down to very low-mass BDs and even toward planetary mass objects. This kind of 
study needs great care because the disk parameters are not direct observables and determining them needs additional assumptions 
like the disk model and dust opacity used in the spectral energy distribution (SED) analysis. Moreover, the {\it Herschel} data 
nowadays is better understood and this knowledge leads to difference with reported fluxes and uncertainties published at 
the early stages of the mission \citep{balog2014}, which will in turn induce discrepancy in the value of parameters used to 
describe the disk structure. Considering these factors, it is not appropriate to analyze the results collected from the 
literatures that reduce and model the {\it Herschel} data in different ways. This may be another reason 
why \citet{olofsson2013} did not find a correlation between the stellar and disk masses in the substellar mass regime
from a comparison between their results and those in the literature. 

In this work, we re-analyzed the {\it Herschel}/PACS data of a sample of 55 BDs and very low mass stars, and fit their 
broadband SEDs using self-consistent radiative transfer models with as few free parameters as possible. We estimated the 
constraints on key disk parameters (i.e., flaring index, scale height and disk mass) through Bayesian analysis. 
We extensively discussed the trends on these parameters as a function of spectral type. Since our analysis is homogeneous 
from data reduction to modeling within a statistically significant sample, we present the first comprehensive study of 
stellar-mass-dependent disk structure down to the planetary mass regime. 

The structure of this paper is as follows: we describe our sample in the following section. 
In Sect. \ref{sec:obs}, we delineate the observations and data reduction. A detailed description of the
modeling is presented in Sect. \ref{sec:modeling}. We discuss our results in Sect. \ref{sec:discussion}, 
followed by a brief summary in Sect. \ref{sec:summary}.

\section{Sample}
\label{sec:sample}
Our sample consists of 55 spectroscopically confirmed targets in nearby star formation regions. 
The spectral types (SpTs) of the sample are in the range from M5.5 to L0. Thus most of the sources have 
substellar masses according to theoretical evolutionary models \citep[e.g.,][]{baraffe2003}.
We do not include any BD candidate lacking spectroscopic confirmation to avoid identification work 
and ambiguity of the results. Evidence for circumstellar dust emission as revealed by thermal IR 
measurements is a crucial criterion of our sample selection. 

The sample of this paper was selected based on the study by \citet{harvey2012a} on one hand and a search 
for the lowest mass objects in the {\it Herschel}/PACS archive on the other hand. \citet{harvey2012a} conducted 
the first comprehensive {\it Herschel}/PACS program to measure far-IR emission from young BDs. We re-perform the 
analysis for the objects in their survey that satisfy the above criteria. For each of these targets, we will provide 
the modeling results that were not reported in their work. In addition, we include the lowest-mass BDs (with SpT 
M8 and later) observed with PACS from the {\it Herschel} archive to probe differences in disk properties between 
high-mass and low-mass BDs. Furthermore, we also add several targets without clear excess at near- and mid-IR, which 
are thought to have experienced significantly inside-out disk evolution, but the cool and outer regions of the disk, if survived, 
are sensitive to {\it Herschel} observations. We note that including all substellar objects from the {\it Herschel}/PACS 
archive in our analysis - although desirable - was not possible due to limited computational power. The sample presented 
here is therefore biased towards very low-mass substellar objects.

Since the SpT is a good proxy for stellar mass in the low stellar mass regime, our sample can be 
generally divided into two groups: (1) a group of 30 early-type brown dwarfs (ETBDs) with spetral type 
earlier than M8, (2) a group of 25 late-type brown dwarfs (LTBDs) with spetral type M8 and later. 
With this kind of division, we can investigate the dependence of the disk structure on the mass 
of the central object in the substellar mass regime, which will yield important clues to understand 
the formation mechanism of BDs and planets. 

Table \ref{tab:sample} summarizes the properties of our sample, including target name, coordinates and SpT.
Our targets are located in eight different star formation regions with ages ranging from
${\sim}1{-}3\,\rm{Myr}$ (e.g., Taurus and Ophiuchus) to ${\sim}10\,\rm{Myr}$ (e.g., TW Hydrae association and Upper Scorpius),
see Table \ref{tab:agedist} for a summary of the cloud ages and distances. The effects of a mixed sample age 
on the results are discussed in detail in Sect. \ref{sec:discussion}. 
Note that for our sample we cannot evaluate the disk frequency of BDs at far-IR due to the incompleteness of sample 
selection and the detection bias in various clouds at different distances to the Earth. 

\begin{landscape}
\begin{table}[!t]
\caption{
\label{tab:sample}
Source summary.}
\renewcommand{\footnoterule}{}
\begin{tabular}{clcccccccccccc}
\hline\hline
ID          & Name   & RA      &    DEC             & ObsIDs  & $F_{\nu}$ 70\,$\mu$m & $F_{\nu}$ 160\,$\mu$m &  SpT  & Cloud  &  Ref  & SD  &  Multiplicity \\[0.15cm]
            &        & (J2000) & (J2000)            &         & [mJy]                & [mJy]                 &       &        &       &     & \\[0.15cm]
\hline
1  & SSSPM1102           & 11 02 09.83 &$-$34 30 35.5 & 1342221849/50   & 9.80 $\pm$ 1.03  & 8.02 $\pm$ 1.10 & M8.5 & TWA & 1, 2   \\
2  & 2M1207              & 12 07 33.47 &$-$39 32 54.0 & 1342202557/58   &7.24 $\pm$ 1.03    & $<$ 17   & M8.0       & TWA & 3, 2  &  & binary \\
3  & 2MJ160859           & 16 08 59.54 &$-$38 56 27.6 & 1342227823/24   & 7.4  $\pm$ 2.0    & $<$ 35  & M8.0 & Lup\,III & 4, 2   \\
4  & GY92 3 (ISO\,032)   & 16 26 21.90 &$-$24 44 39.8 & 1342238816/17   &49.75 $\pm$ 5.58   & $<$ 647  & M8.0 & Oph & 5, 6   \\
5  & GY92 264            & 16 27 26.58 &$-$24 25 54.4 & 1342238816/17   &33.26 $\pm$ 5.45   & $<$ 1489 & M8.0 & Oph & 5, 6   \\
6  & GY92 310            & 16 27 38.63 &$-$24 38 39.2 & 1342238816/17   &120.28 $\pm$ 5.45  & $<$ 1032 & M8.5 & Oph & 7, 6 & Y1  \\
7  & CFHTWIR96           & 16 27 40.84 &$-$24 29 00.7 & 1342227841/42   &$<$ 193            & $<$ 1192& M8.25 & Oph & 8, 2   \\
8  & OTS\,44             & 11 10 09.34 &$-$76 32 17.9 & 1342238816/17   &4.63 $\pm$ 1.03    & $<$ 211  & M9.5 & Cha\,I    & 9, 10  \\
9  & 2MJ111145           & 11 11 45.34 &$-$76 36 50.5 & 1342223476/77   &2.84 $\pm$ 1.03    & $<$ 33   & M8.0 & Cha\,I    & 11, 2  \\
10 & J125758.7-770120    & 12 57 58.70 &$-$77 01 19.5 & 1342212708/09   &$-$                & $<$ 243  & M9.0 & Cha\,II   & 12, 13 \\
11 & KPNO 6              & 04 30 07.24 &+26 08 20.8 & 1342227011/12   &$<$ 4              & $<$ 47  & M9.0  & Tau       & 14, 2  \\
12 & KPNO 7              & 04 30 57.19 &+25 56 39.5 & 1342227999/8000 &2.61 $\pm$ 1.03    & $<$ 69  & M8.25 & Tau & 14, 2  \\
13 & J04574903+3015195*   & 04 57 49.03 &+30 15 19.5 & 1342243478/79   &$<$ 8             & $<$ 14   & M9.25 & Tau       & 15, 16 \\
14 & J04354526+2737130*   & 04 35 45.26 &+27 37 13.1 & 1342243466/67   &$<$ 11            & $<$ 14   & M9.2  & Tau       & 17, 16 \\
15 & J04334291+2526470*   & 04 33 42.92 &+25 26 47.0 & 1342243084/85   &$<$ 7             & $<$ 20   & M8.75 & Tau & 17, 16 \\
16 & J04335245+2612548   & 04 33 52.46 &+26 12 54.9 & 1342243080/81   &8.00 $\pm$ 2.32   & $<$ 26   & M8.5  & Tau       & 18, 16 & Y1 \\
17 & J04290068+2755033   & 04 29 00.68 &+27 55 03.4 & 1342243066/67   & $<$ 9            & $<$ 36   & M8.25 & Tau & 17, 16 \\
18 & KPNO 4*       & 04 27 28.00 &+26 12 05.3 & 1342242064/65   & $<$ 9            & $<$ 13   & L0    & Tau       & 15, 16 \\
19 & J04215450+2652315*  & 04 21 54.51 &+26 52 31.5 & 1342242033/34   & $<$ 7            & $<$ 23   & M8.5  & Tau       & 17, 16 \\
20 & J04263055+2443558   & 04 26 30.55 &+24 43 55.9 & 1342242023/24   & 8.00 $\pm$ 2.22  & $<$ 33   & M8.75 & Tau & 17, 16 & Y1 \\
21 & J04274538+2357243*   & 04 27 45.38 &+23 57 24.3 & 1342242021/22   & $<$ 9            & $<$ 23   & M8.25 & Tau & 17, 16 \\
22 & J04302365+2359129   & 04 30 23.66 &+23 59 13.0 & 1342242009/10   & $<$ 7            & $<$ 13   & M8.5 & Tau        & 17, 16 \\
23 & J04311907+2335047*   & 04 31 19.07 &+23 35 04.7 & 1342242007/08   & $<$ 8            & $<$ 15   & M8.0 & Tau        & 19, 16 \\
24 & KPNO 9        & 04 35 51.43 &+22 49 12.0 & 1342241979/80   & $<$ 12           & $<$ 16   & M8.5 & Tau        & 20, 16 \\
25 & J04361030+2159364   & 04 36 10.31 &+21 59 36.5 & 1342241975/76   & $<$ 8            & $<$ 16   & M8.5 & Tau        & 17, 16 \\
26 & J04325119+1730092*   & 04 32 51.20 &+17 30 09.2 & 1342241957/58   & $<$ 9            & $<$ 19   & M8.25 & Tau & 21, 16 \\
27 & KPNO 1*        & 04 15 14.71 &+28 00 09.6 & 1342241914/15   & $<$ 6            & $<$ 19   & M8.75 & Tau & 15, 16 \\
28 & KPNO 12       & 04 19 01.27 &+28 02 48.7 & 1342241888/89   & $<$ 6            & $<$ 49   & M9.25& Tau        & 15, 16 \\
29 & J04221332+1934392   & 04 22 13.32 &+19 34 39.2 & 1342241878/79   & $<$ 10           & $<$ 24   & M8.0 & Tau        & 17, 16  &   & binary \\
30 & J04414489+2301513   & 04 41 44.90 &+23 01 51.4 & 1342240750/51   & $<$ 12           & $<$ 17   & M8.5 & Tau        & 17, 16 & Y1 & binary \\[0.15cm]
\hline \myrule
31 &   ISO\,138          & 11 08 18.51 &$-$77 30 40.8 & 1342218699/700  & 4.52 $\pm$ 1.02   & $<$ 89   & M6.5  & Cha\,I   & 22, 2 \\
32 &    ISO\,217  & 11 09 52.16 &$-$76 39 12.8 & 1342223474/75   & 10.47 $\pm$ 1.10  & $<$ 507  & M6.25  & Cha\,I  & 23, 2 \\
33 &     CHAHA6          & 11 08 39.52 &$-$77 34 16.7 & 1342223486/87   & 14.50 $\pm$ 1.10  & $<$ 54   & M5.75  & Cha\,I  & 23, 2 \\
34 &  2MJ110703          & 11 07 03.69 &$-$77 24 30.7 & 1342223480/81   & 28.48 $\pm$ 1.02  & $<$ 447  & M7.5  & Cha\,I   & 24, 2 & Y1, Y2 \\
35 &     CHAHA9          & 11 07 18.61 &$-$77 32 51.7 & 1342223482/83   & 15.30 $\pm$ 1.01  & $<$ 220  & M5.5  & Cha\,I   & 23, 2 & Y1 \\
36 &   ISO\,252          & 11 10 41.42 &$-$77 20 48.1 & 1342223478/79   & 8.81  $\pm$ 1.10  & $<$ 122  & M6.0  & Cha\,I   & 23, 2 \\
37 &   ISO\,165          & 11 08 54.97 &$-$76 32 41.1 & 1342223470/71   & 45.46 $\pm$ 1.02  & 31.89 $\pm$ 7.03 & M5.5 & Cha\,I & 23, 2 \\
38 &  CHAII1258          & 12 58 06.76 &$-$77 09 09.5 & 1342224206/07   & 5.06  $\pm$ 1.01  & $<$ 116  & M6.0  & Cha\,II  & 25, 2 \\
39 &  CHAII1308          & 13 08 27.14 &$-$77 43 23.3 & 1342224204/05   & 6.43  $\pm$ 1.01  & $<$ 37   & M6.0  & Cha\,II  & 25, 2 \\
40 &  CFHTTAU12          & 04 33 09.46 &+22 46 48.7 & 1342227013/14   & $<$ 7             & $<$ 54   & M6.5  & Tau      & 17, 2 \\
\hline
\end{tabular}
\end{table}
\end{landscape}

\addtocounter{table}{-1}
\begin{landscape}
\begin{table}[!t]
\caption{continued.}
\renewcommand{\footnoterule}{}
\begin{tabular}{clcccccccccccc}
\hline\hline
ID          & Name   & RA      &  DEC                               & ObsIDs  & $F_{\nu}$ 70\,$\mu$m & $F_{\nu}$ 160\,$\mu$m &  SpT  & Cloud  &  Ref  & SD & Multiplicity \\[0.15cm]
            &        & (J2000) & (J2000)                            &         & [mJy]                & [mJy]                 &       &        &       &      \\[0.15cm]
\hline
41 &   CFHTTAU9          & 04 24 26.46 &   +26 49 50.4 & 1342227059/60   & 10.64 $\pm$ 1.09  & $<$ 26   & M6.25  & Tau  & 17, 2 \\
42 & SORI053825          & 05 38 25.44 & $-$02 42 41.3 & 1342226721/22   & $<$ 12            & $<$ 22   & M7.0  & $\sigma$ Ori& 26, 2 \\
43 & OPHALL1622          & 16 22 44.94 & $-$23 17 13.4 & 1342227839/40   & 5.75  $\pm$ 1.02  & $<$ 160  & M6.0  & Oph      & 25, 2  & Y1 \\
44 &  2MJ160737          & 16 07 37.73 & $-$39 21 38.8 & 1342227813/14   & 16.15 $\pm$ 1.02  & 8.54 $\pm$ 1.61   &  M5.75 & Lup\,III & 4, 2 \\
45 & DENIS\,1603         & 16 03 34.71 & $-$18 29 30.4 & 1342227831/32   & 3.60  $\pm$ 1.01  & $<$ 22   & M5.5  & Upper Sco & 27, 2 \\
46 &    USCO112          & 16 00 26.70 & $-$20 56 31.6 & 1342227140/41   & $<$ 5             & $<$ 12   & M5.5  & Upper Sco & 28, 2 \\
47 &    USCO128          & 15 59 11.36 & $-$23 38 00.2 & 1342227134/35   & 5.69  $\pm$ 1.02  & $<$ 25   & M7.0  & Upper Sco & 28, 2 & Y2 \\
48 &     USCO55          & 16 02 45.60 & $-$23 04 50.0 & 1342227136/37   & 13.26 $\pm$ 1.02 & $<$ 11   & M5.5  & Upper Sco  & 28, 2 & Y2 \\
49 &  USD155556          & 15 55 56.01 & $-$20 45 18.7 & 1342227138/39   & 15.92 $\pm$ 1.01 & $<$ 11   & M6.5  & Upper Sco  & 27, 2 & Y1, Y2 \\
50 &  USD155601          & 15 56 01.04 & $-$23 38 08.1 & 1342227132/33   & 7.52  $\pm$ 1.01 & $<$ 37   & M6.5  & Upper Sco  & 27, 2 & Y2 \\
51 &  USD160603          & 16 06 03.91 & $-$20 56 44.4 & 1342227142/43   & 3.49  $\pm$ 1.01 & $<$ 24   & M7.5  & Upper Sco  & 27, 2 \\
52 &  USD160958          & 16 09 58.53 & $-$23 45 18.6 & 1342227146/47   & 6.07  $\pm$ 1.01 & $<$ 18   & M6.5  & Upper Sco  & 27, 2 \\
53 &  USD161005          & 16 10 05.42 & $-$19 19 36.3 & 1342227829/30   & 5.37  $\pm$ 1.01 & 8.80 $\pm$ 1.61  & M7.0 & Upper Sco & 27, 2 \\
54 &  USD161833          & 16 18 33.18 & $-$25 17 50.5 & 1342227827/28   & 10.42 $\pm$ 1.02 & $<$ 93   & M6.0  & Upper Sco  & 27, 2 & & binary \\
55 &  USD161939          & 16 19 39.76 & $-$21 45 35.0 & 1342227835/36   & 15.88 $\pm$ 1.01 & $<$ 42   & M7.0  & Upper Sco  & 27, 2 & & binary \\
\hline
\end{tabular}
\tablefoot{(1) ``$\ast$'' marks sources without clear excess emission at near- and mid-IR wavelengths.
(2) The horizontal line divides the sample into a group of late-type BDs (M8 and later, top) and a group of early-type 
BDs (earlier than M8, bottom). (3) The ObsIDs contain both the scan and cross scan ID numbers of the observations.
Taking SSSPM1102 as an example, 1342221849 is the scan ID and 1342221850 is the cross-scan ID. All ObsIDs follow the same convention.
(4) First reference listed is for spectral type. Further information (e.g., near- and mid-IR photometry) can be found in the second 
reference and references therein. (5) The column ``SD'' marks the objects for which the re-measured PACS photometry are significantly 
different from the literature values at $70\,\mu{\rm{m}}$ (Y1) and $160\,\mu{\rm{m}}$ (Y2) respectively, see Sect. \ref{sec:obs_trend}. \\
{\bf References.} (1) \citet{scholz2005}; (2) \citet{harvey2012a}; (3) \citet{gizis2002}; (4) \citet{allen2007};
(5) \citet{wilking2005}; (6) \citet{alves2013}; (7) \citet{wilking1999}; (8) \citet{alves2010}; (9) \citet{luhman2004a};
(10) \citet{joergens2013}; (11) \citet{luhman2007}; (12) \citet{spezzi2008}; (13) \citet{spezzi2013}; 
(14) \citet{guieu2007}; (15) \citet{canty2013}; (16) \citet{bulger2014}; (17) \citet{luhman2006a};
(18) \citet{luhman2006b}; (19) \citet{slesnick2006}; (20) \citet{briceno2002}; (21) \citet{luhman2009};
(22) \citet{luhman2004b}; (23) \citet{damjanov2007}; (24) \citet{luhman2008}; (25) \citet{gully2011};
(26) \citet{rigliaco2011}; (27)  \citet{martin2004}; (28) \citet{ardila2000}.
}
\end{table}
\end{landscape}

\begin{table}[!t]
\begin{center}
\caption{Cloud age and distance assumed in this work.}
\begin{tabular}{lcccc}
\hline\hline
Cloud          &     Distance          &      Age    &  Reference \\ 
               &      (pc)             &     (Myr)   &            \\  
 \hline
Ophiuchus (Oph)&  $130$       &       $1-3$       &   1, 2        \\
Taurus  (Tau)       &  $142.5$     &       $1-3$       &   3, 4        \\
Chamaeleon I (Cha I)    &  $162.5$     &       $1-3$       &   1, 5     \\
Chamaeleon II (Cha II)  &  $178$       &       $1-3$       &   1, 6        \\
Lupus III (Lup III)     &  $200$       &       ${\sim}5$   &   1, 7       \\
$\sigma$ Orionis ($\sigma$ Ori)   &  $360$  &       ${\sim}3$   &   1, 8        \\
Upper Scorpius (Upper Sco)  &  $145$       &       ${\sim}11$  &   1, 9        \\
TW Hydrae association (TWA) &  $55$       & $8-10$      &   10, 11        \\
\hline
\end{tabular}
\label{tab:agedist}
\end{center}
{\bf References.} (1) \citet{reipurth2008a}; (2) \citet{wilking2005};
(3) \citet{reipurth2008b}; (4) \citet{briceno2002}; (5) \citet{luhman2007};
(6) \citet{spezzi2008}; (7) \citet{comeron2009}; (8) \citet{sherry2004}; 
(9) \citet{pecaut2012}; (10) \citet{ducourant2014}; (11) \citet{barrado2006}. 
\end{table}

\section{Observations and data reduction}
\label{sec:obs}

All the targets have been observed by the Photodetector Array Camera and Spectrograph \citep[PACS;][]{poglitsch2010} 
on board {\it Herschel} with various programs \citep{andre2010,harvey2012a,alves2013,bulger2014}. The ObsIDs for 
all the observations are listed in Table \ref{tab:sample}. The majority of the observations were performed 
in the mini-scan map mode. The only exceptions are those for GY92\,3, GY92\,264, GY92\,310 and J125758.7-770120. 
They were observed in the standard PACS scan map mode. Parallel mode observations were not considered.
We took the data from the {\it Herschel} Science Archive at the processing stage of Level 1, Version 12.1 of 
the standard pipeline (SPG = Standard Product Generation) and further processed the data by producing maps and 
by conducting photometry. 

\subsection{Mapping}
The Level 1 data from the archive are fully calibrated \citep{balog2014}. Only the last steps to Level 2 and 2.5 are missing, which 
consist of deglitching, signal drift suppression and mapping to convert the detector data into a map that 
superimposes the detector signals for any given position on the sky covered during the observation.

Since we are only interested in the point sources, the highpass filtering 
algorithm was applied to get rid of the background and detector signal drifts. 
This approach only conserves the source flux, if an object mask indicates those 
parts of the data that are not considered as background. For very faint 
objects like BDs, the mask produced by the SPG is insufficient. 
Therefore, we produced new and adequate source masks from Level 2.5 SPG 
maps (scan and cross-scan pairs are coadded) based on a signal-to-noise criterion.

After deglitching the data as done by the SPG, we produced maps using the 
\texttt{photProject} task. It converts the flux distribution as measured by 
the fixed detector pixel grid into a rebinned map, where the flux is redistributed on a 
freely selectable and virtual pixel grid. 
Since the map pixels are usually smaller, distorted and rotated relative to the detector 
pixels, the flux distribution must and can be done with great care. There is a 
``drizzling'' algorithm, originally invented for the {\it HST} that allows modifying 
crucial parameters when doing this \citep{fruchter2002}.

The SPG uses a so-called drop size that is relatively large. This is a scaling 
factor of the detector pixel sizes that is applied before projecting them on 
the final map grid. This leads to a redistribution of the source flux on larger 
areas of the final map. This produces prettier, i.e. smoother images due to the 
convolution of neighbouring detector pixel signals. But this also averages out 
noise, which leads to an underestimate of the real detector noise. Reducing the 
drop size can minimize this effect. The images look sharper, but also noisier. 
However, the resulting noise properties are much closer to the reality than 
what one gets with large drop sizes. Therefore, the subsequent photometry is 
also more accurate. We adopted 0.1 as the drop size for all maps. The pixel sizes 
were chosen to be $1^{\prime\prime}$ at $70\,\mu{\rm{m}}$ and $2^{\prime\prime}$ 
at $160\,\mu{\rm{m}}$.

Finally, we combine the map pairs of a so-called scan and a cross scan, corresponding 
to the Level 2.5 stage of the SPG.

\begin{figure}[!t]
\includegraphics[width=0.45\textwidth]{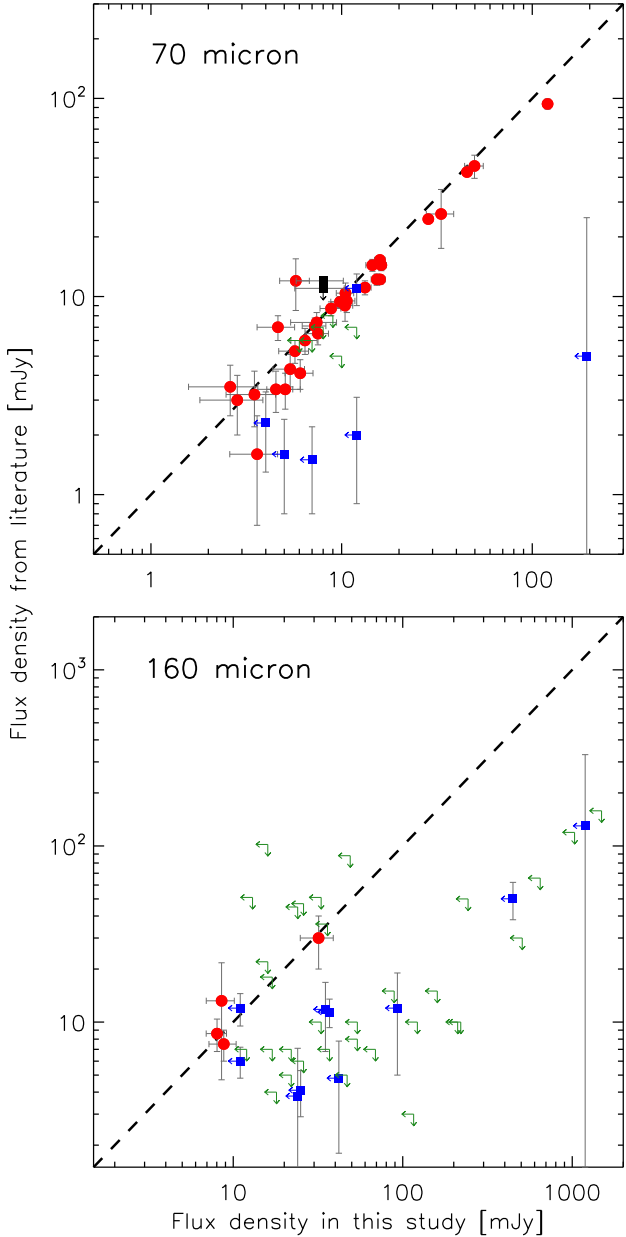}
\caption{Our {\it Herschel}/PACS flux densities at $70\,\mu{\rm{m}}$ (upper panel) and $160\,\mu{\rm{m}}$ (lower panel) 
         as compared to previous results. The red dots depict real detections identified by both our and 
         previous works, whereas green marks consisting of leftward and downward arrows symbolize the upper 
         limits. The black squares with downward arrows and blue squares with leftward arrows show detections by our 
         study and previous works, respectively.}
\label{fig:compare}
\end{figure}

\subsection{Photometry}
Since the sources are very faint and the vicinities around the targets of 
interest are full of other and brighter sources, a simple aperture photometry 
procedure is not sufficient. Therefore, we applied a HIPE built-in source extraction 
and photometry tool called \texttt{Sussextractor}. It is based on Bayesian statistics and 
delivers the most likely result for the target position and the integrated source flux 
by fitting 2D Gaussians to the detected sources \citep{savage2007}.

A Gaussian source model is not adequate for the more complicated PSF shape 
of {\it Herschel}/PACS point sources. Only the core of the PACS PSF is close to 
Gaussian, but it also contains a trilobal shape at a level of a few percent of 
the peak intensity superimposed by an extended and slowly declining halo. 
Therefore, a correction factor has to be applied to the photometry derived by 
Sussextractor.

For this purpose, we reprocessed standard star data of alpha Boo (a PACS prime standard) 
in the same way as the BDs and then applied the \texttt{Sussextractor} algorithm and 
compared the results with the ones derived from a standardized aperture 
photometry procedure including an aperture correction that was derived during 
the PACS flux calibration program. The multiplicative correction factors 
turned out to be 1.513 at 70\,$\mu$m and 1.427 at 160\,$\mu$m. The photometry 
errors given by the Sussextractor tool were corrected in the same way and regarded as the 
photometric errors. Additional systematic uncertainties were not considered.
For non-detections, upper limits have been obtained by a statistical analysis of a 
series of aperture photometry measurements on the background as explained in \citet{balog2014}. 
This approach avoids underestimating the noise obtained from the map pixel statistics 
that is affected by correlation artefacts introduced by the drizzling algorithm.

\begin{figure}[!t]
\includegraphics[width=0.45\textwidth]{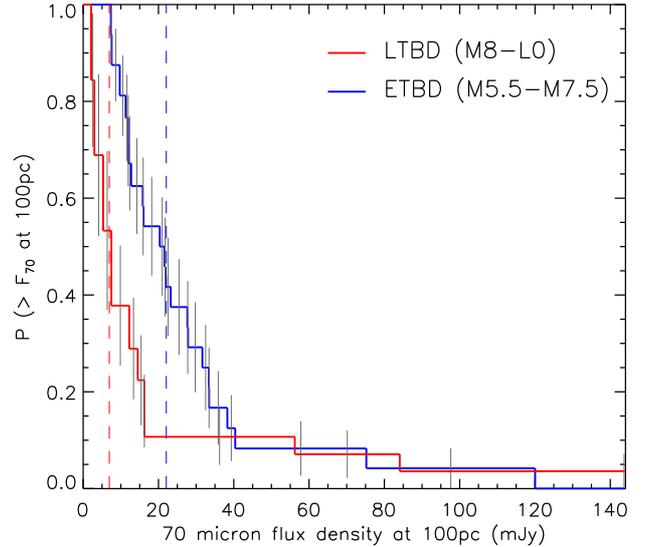}
   \caption{The cumulative distribution of the $70\,\mu{\rm m}$ flux density for early-type brown dwarfs (ETBDs, blue) and 
            late-type brown dwarfs (LTBDs, red). All flux densities are scaled assuming a distance of 100\,pc. 
            The median values are indicated as vertical dashed lines.}
\label{fig:medflux}
\end{figure}

\subsection{General trends of the observational data}
\label{sec:obs_trend}
The measured flux densities at PACS bands are summarized in Table \ref{tab:sample}, whereas Figure \ref{fig:compare} 
shows a comparison between our results and previously published photometry. It is clear that most 
detections at $70\,\mu{\rm{m}}$ by other studies are confirmed with our reanalysis of the 
data. In particular, our flux measurements of the detected sources are in agreement with previous 
results (see the red dots in the upper panel of Figure \ref{fig:compare}). However, we obtained less than 
one half of $160\,\mu{\rm{m}}$ detections that are reported before. This is probably because {\it Herschel} 
$160\,\mu{\rm{m}}$ images of faint disks are easily contaminated with the extended emission from the 
background, and the effects of using different data reduction algorithms on the results are significant in 
this case. In Table \ref{tab:sample}, we mark two kinds of objects for which the re-analyzed PACS photometry 
are significantly different from the literature values: (1) the difference in the flux density between 
our and previous results is larger than $3\,\sigma$, where $\sigma$ refers to the RMS level for each object given
here; (2) firm detections reported before, but upper limits deduced by us, and vice versa.
Since the constraints on the disk structure, especially the structure of the outer and 
cool regions of the disk, are sensitive to far-IR data points \citep[e.g.,][]{harvey2012a, harvey2012b, olofsson2013}, 
the discrepancies in the PACS photometry between our and previous results will result in different model explanations of
the data. Therefore, our homogeneous analysis of the {\it Herschel} data and subsequent self-consistent modeling have 
the potential to retrieve the hidden statistical trends of disk properties in the substellar mass regime.

We generated the cumulative distributions $f(\ge F_{\nu})$ of PACS photometry, using the Kaplan-Meier product-limit
estimator to properly account for censored data sets, i.e., upper limits on $F_{\nu}$ \citep{feigelson1985}. This 
was done to examine any statistical difference in the far-IR emission between the two groups differing in terms 
of SpT. All flux densities are scaled by assuming a distance of 100\,pc. The results are shown in Figure \ref{fig:medflux}. 
The LTBD sample (red line) has systematically lower flux densities than those of the ETBD group (blue line), which
is consistent with the finding by \citet{liu2015} that the amount of {\it Herschel} far-IR emission correlates well 
with the SpTs of the host BDs. From the distribution, the median flux densities are estimated, which 
we show as the vertical dashed lines. The median $70\,\mu{\rm{m}}$ flux density of the ETBD group is $F_{70}=22\,\rm{mJy}$,
which is at least three times larger than that of the LTBDs, i.e., $F_{70}<7\,\rm{mJy}$. We consider the latter 
case as an upper limit because more than half of the sources are undetected at this wavelength \citep{mathews2013}. 
Due to the same reason, we do not perform the Kaplan-Meier product-limit estimation for the $160\,\mu{\rm{m}}$ data. 

From a theoretical point of view, disks with a larger flaring index intercept a larger portion of the central star's 
radiation due to a larger flaring angle and consequently re-emit more IR flux \citep[e.g.,][]{chiang1997}. The 
systematically lower flux density at far-IR wavelengths of the LTBDs implies that disks around lower mass stars 
are generally less flared. \citet{bulger2014} investigated the effect of individual disk parameter on the 
appearance of the SED, and their results indicate that the flaring index and scale height work together to 
determine the emission level at {\it Herschel}/PACS bands. Moreover, the observational trends of far-IR flux densities may also 
help us to consider whether there is any evidence for differences in disk mass with different stellar masses, since the 
{\it Herschel}/PACS photometry can provide a rough mass estimation of BD disks \citep[e.g.,][]{harvey2012a, spezzi2013, liu2015}. 
Nevertheless, interpreting the observational trends presented here needs detailed SED analysis, in which efforts 
of reducing model degeneracy should be made as far as possible given that the data available are limited.

\section{SED Modeling}
\label{sec:modeling}
All the targets have been observed over a broad range of wavelengths. We constructed the broadband SEDs of our 
sources by using our {\it Herschel}/PACS flux measurements and adding ancillary data at optical and mid-IR 
wavelengths, from SDSS catalog \citep{ahn2012}, DENIS survey \citep{denis2005}, 2MASS \citep{cutri2003}, {\it Spitzer} 
and WISE catalogs \citep[e.g.,][]{cutri2012}. We also took into account the upper limits of flux measurements 
in the (sub-)millimeter regime available for a few of our sources \citep[e.g.,][]{mohanty2013, broekhoven2014}.
We derived the (sub)stellar properties of each target from thoroughly modeling the photosphere. For the 46 targets 
that show IR excess produced by circumstellar dust, we conducted detailed SED analysis using the radiative transfer 
code \texttt{MC3D} developed by \citet{wolf2003a} in order to characterize the structure of their surrounding disks. 

\subsection{Determination of the (sub)stellar properties}
\label{sec:stellar}
We performed a thorough modeling of the photosphere of each object by fitting the BT-Settl 
models \citep{allardh2012} with a $\chi^2$ minimization as well as Bayesian statistics \citep{bayo2008, bayo2014}.
The BT-Settl models incorporate a sophisticated treatment of photospheric dust, which is 
likely to affect the atmospheres and correspondingly the radiation field of cool objects like 
BDs in our sample. The broad-band photometry for the photospheric part of the SED are taken into
account in the fitting procedure. A similar analysis was done by \citet{joergens2013} for their 
case study of OTS\,44, a BD at the planetary mass border.

For objects in star-forming regions, we use the standard distances with errors to the cloud as given by 
the Handbook of Star formation \citep{reipurth2008a, reipurth2008b}, while the kinematic distances for objects 
in loose associations are adopted \citep[e.g.,][]{ducourant2014}. The extinction is treated as a free parameter. 
For objects in star-forming regions, we use as starting point upper limits for the extinction based on extinction 
maps from \citet{kainulainen2009} and \citet{cambresy1997}. For objects in loose associations, we consider 
an upper limit of 0.1\,mag, and for $\sigma$\,Ori, we convert the standard E(B-V) value to extinction 
$A_{V}$ assuming $R_V=3.1$ since this is the basic assumption for the extinction law used in VOSA \citep{bayo2008}.


The derived best-fit ($\chi^2$) and most probable values for the effective temperatures, luminosities, and extinctions 
are in good agreement with each other for ${\sim}90\%$ of the objects. For a few cases, we only get moderate 
fitting results: (i) for object 21 (J04274538+2357243), the extinction $A_{V}$ is loosely constrained, we therefore 
took the value that minimizes the $\chi^2$. 
(ii) for object 34 (2MJ110703), there is no data point bluer than 2MASS/J band. In this case, we fixed  
$A_{V}$ to 15\,mag which is lower than the value of 16.5\,mag estimated in \citet{furlan2009}, but is 
in agreement with the one derived by \citet{luhman2008}. 
(iii) object 36 (ISO\,252) for which the best and second best fits are indistinguishable in terms of $\chi^2$ and 
the Bayesian analysis shows that it is hard to break the degeneracy between $T_{\rm{eff}}$ and $A_{V}$.  
We took the best-fit parameter set for this object.
The derived effective temperatures, luminosities and extinctions are summarized in Table \ref{tab:param}. 

\subsection{Disk model}
\label{sec:model}
\noindent{\it Dust distribution:}\hspace*{2mm}
We employed the standard flared disk model with well-mixed gas and dust, which has been successfully used to explain 
the observed SEDs of a large sample of young stellar objects and BDs \citep[e.g.,][]{wolfp2003, sauter2009, harvey2012a, joergens2013, liu2015}.
The structure of the dust density is assumed with a Gaussian vertical profile
\begin{equation}
\rho_{\rm{dust}}=\rho_{0}\left(\frac{R_{*}}{\varpi}\right)^{\alpha}\exp\left[-\frac{1}{2}\left(\frac{z}{h(\varpi)}\right)^2\right], \\
\label{dust_density}
\end{equation}
and the surface density is described as a power-law function  \\
\begin{equation}
\Sigma(\varpi)=\Sigma_{0}\left(\frac{R_{*}}{\varpi}\right)^p,
\end{equation}
where $\varpi$ is the radial distance from the central star measured in the disk midplane, and $h(\varpi)$ is the scale 
height of the disk. The disk extends from an inner radius $R_{\rm{in}}$ to an outer radius $R_{\rm{out}}$. 
To the best of our knowledge, among our sample, there are 5 objects that have been identified as binary systems so far.
They are 2M1207 \citep[$a{\sim}55\,\rm{AU}$,][]{chauvin2004}, J04221332+1934392 \citep[$a{\sim}7\,\rm{AU}$,][]{todorov2014},
J04414489+2301513 \citep[$a{\sim}15\,\rm{AU}$,][]{todorov2014}, USD161833 \citep[$a{\sim}134\,\rm{AU}$,][]{bouy2006}
and USD161939 \citep[$a{\sim}26\,\rm{AU}$,][]{bouy2006}, where $a$ refers to the separation within the system.
The disks around individual components in binary systems are expected to have truncation radii of the 
order of $(0.3-0.5)a$ \citep{papaloizou1977}. We adopted $0.5\,a$ as the disk outer radii for 
2M1207, USD161833 and USD161939. For the close pairs ($a{\lesssim15\,\rm{AU}}$, J04221332+1934392 and J04414489+2301513),
dynamical simulations of star-disk interactions suggest that individual disks are unlikely to survive \citep[e.g.,][]{artymowicz1994}. 
Disk modeling is complicated in those close multiple systems. For simplicity, we assume that the emission is associated with 
circumbinary disks of $100\,\rm{AU}$ in size. For other objects, we fix $R_{\rm{out}}=100\,\rm{AU}$ in the modeling, because 
the choice of this parameter value makes essentially no difference to the synthetic SEDs in the simulated wavelength 
range \citep{harvey2012a}. The scale height follows the power law distribution
\begin{equation}
h(\varpi) = H_{100}\left(\frac{\varpi}{100\,\rm{AU}}\right)^\beta,\\
\end{equation}
with the exponent $\beta$ characterizing the degree of flaring and $H_{100}$ representing the scale height at a distance of 
$100\,\rm{AU}$ from the central star. The indices $\alpha$, $p$, and $\beta$ are codependent through $p=\alpha-\beta$. 
We fix $p=1$ that is the typical value found for T Tauri disks in the sub-millimeter \citep[e.g.,][]{,isella2009,guilloteau2011},
since only spatially resolved data can place constraints on this parameter \citep[e.g.,][]{ricci2013, ricci2014}. 
~~~~~\\ ~~~~~\\
\noindent{\it Dust properties:}\hspace*{2mm}
We assume the dust grains to be a homogeneous mixture of 75\% amorphous silicate and 25\% carbon with 
a mean density of $\rho_{\rm{grain}} = 2.5\,{\rm g\,cm^{-3}}$ and the complex refractive indices given
by \citet{jager1994}, \citet{dorschner1995} and \citet{jager1998}. Porous grains are not considered 
because the fluxes at wavelengths beyond ${\sim}2\,\mu{\rm{m}}$ are almost independent of the degree of 
grain porosity in low mass disks, as shown by \citet{kirchschlager2014}.
The grain size distribution is given by the standard power law ${\rm d}n(a)\propto{a^{-3.5}} {\rm d}a$ with 
minimum and maximum grain sizes $a_{\rm{min}}=0.1\,\mu{\rm m}$ and $a_{\rm{max}}=100\,\mu{\rm m}$, respectively.
The choice of the minimum value for the grain size, $a_{\rm{min}}$, ensures that its exact value has a negligible impact on the 
synthetic SEDs. Since there is no information about the maximum grain sizes of our target disks, as provided e.g., by the 
(sub-)millimeter spectral index, we adopt $a_{\rm{max}}=100\,\mu{\rm{m}}$. The {\it Herschel}/PACS far-IR observations are 
sensitive to dust grains with this assumed sizes. Strong grain growth up to millimeter sizes as detected in some BD disks 
\citep[e.g.,][]{ricci2012, ricci2013, broekhoven2014, ricci2014} would remain undetected in our data and could affect the 
disk mass. Our prescription for the dust properties is identical to those used in \citet{liu2015}.  

~~~~~\\ ~~~~~\\
\noindent{\it Heating sources:}\hspace*{2mm}
The disk is assumed to be passively heated by stellar irradiation \citep[e.g.,][]{chiang1997, dullemond2001}. 
Other heating sources, such as the viscous accretion, are not taken into account as this would only introduce 
further free parameters without any qualitative constraints from the modeling. 
We use the BT-Settl atmosphere models \citep{allardh2012} as incident flux with parameters as 
constrained in Sect. \ref{sec:stellar}. The radiative transfer problem is solved self-consistently 
considering 100 wavelengths, which are logarithmically distributed in the range 
of [$0.05\,\mu{\rm{m}}$, $2000\,\mu{\rm{m}}$].

\begin{table*}[!t]
\begin{center}
\caption{
\label{tab:param}
Photospheric and best-fit/most-probable disk parameters from this work.
}
\renewcommand{\footnoterule}{}
\begin{tabular}{ccllcccccccc}
\hline\hline
ID     & $T_{\rm{eff}}$ & $L_*$                  & $A_V$ & $R_{\rm in}$ & $M_{\rm disk}\,(B)$        &  $M_{\rm disk}\,(P)$   & $\beta\,(B)$  & $\beta\,(P)$ & $H_{\rm 100}\,(B)$ & $H_{\rm 100}\,(P)$  & $i$ \\[0.15cm]
       &    [K]         & [10$^{-3}\,L_{\odot}$] & [mag] & [AU]         & [10$^{-5}\,\rm M_{\odot}$] &  [10$^{-5}\,\rm M_{\odot}$]  &    &    &  [AU]  & [AU] & [$^{\circ}$] \\ 
\hline
1      & 2400 & 0.355 $\pm$ 0.106  & 0.1  $\pm$  0.01 &  0.004 & 1.0  & 1.0   & 1.10  & 1.10  & 20  & 20  & 60  \\
2      & 2500 & 2.367 $\pm$ 0.336  & 0.1  $\pm$  0.01 &  0.025 & 0.1  & 0.1   & 1.175 & 1.175 & 14  & 14  & 45  \\
3      & 2100 & 6.139 $\pm$ 1.852  & 0.26 $\pm$  0.26 &  0.038 & 3.16 & 1.0   & 1.10  & 1.10  & 16  & 16  & 15  \\
4      & 2800 & 64.2  $\pm$ 10.49  & 3.6  $\pm$  0.15 &  0.041 & 3.16 & 3.16  & 1.10  & 1.10  & 16  & 16  & 30  \\
5      & 2800 & 23.18 $\pm$ 4.020  & 2.28 $\pm$  0.18 &  0.025 & 10.0 & 10.0  & 1.10  & 1.125 & 20  & 18  & 30  \\
6      & 2800 & 81.15 $\pm$ 12.94  & 7.2  $\pm$  0.25 &  0.317 & 10.0 & 10.0  & 1.125 & 1.125 & 14  & 14  & 60  \\
7      & 2600 & 379.1 $\pm$ 58.95  & 3.94 $\pm$  0.98 &  0.015 & 0.1  & 0.32  & 1.15  & 1.15  & 20  & 20  & 15  \\
8      & 1700 & 2.4   $\pm$ 0.54   & 2.6  $\pm$  0.20 &  0.023 & 3.25 & 3.16  & 1.31  & 1.30  & 15.6& 16  & 60  \\
9      & 2800 & 5.459 $\pm$ 0.352  & 3.0  $\pm$  0.19 &  0.038 & 3.16 & 3.16  & 1.20  & 1.175 &  8  & 10  & 15  \\
10     & 2500 & 6.242 $\pm$ 0.325  & 4.8  $\pm$  0.22 &  0.005 & 0.1  & 0.1   & 1.025 & 1.05  & 18  & 18  & 60  \\
11     & 2600 & 3.54  $\pm$ 0.192  & 1.35 $\pm$  0.1  &  0.005 & 1.0  & 0.32  & 1.10  & 1.10  & 8   & 10  & 30  \\
12     & 2600 & 5.603 $\pm$ 0.376  & 1.41 $\pm$  0.07 &  0.038 & 1.0  & 1.0   & 1.10  & 1.10  & 8   & 8   & 60  \\
13     & 2200 & 1.221 $\pm$ 0.080  & 0.2  $\pm$  0.03 & --     & --   & --    & --    & --    & --  & --  & --  \\
14     & 2400 & 3.22  $\pm$ 0.152  & 1.27 $\pm$  0.07 & --     & --   & --    & --    & --    & --  & --  & --  \\
15     & 2500 & 5.591 $\pm$ 0.237  & 1.89 $\pm$  0.1  & --     & --   & --    & --    & --    & --  & --  & --  \\
16     & 2600 & 4.078 $\pm$ 0.177  & 3.91 $\pm$  0.23 &  0.010 & 100  & 100   & 1.10  & 1.075 & 20  & 20  & 45  \\
17     & 2700 & 10.43 $\pm$ 0.429  & 1.71 $\pm$  0.1  &  0.051 & 3.16 & 1.0   & 1.10  & 1.125 & 8   & 10  & 45  \\
18     & 1800 & 3.045 $\pm$ 0.185  & 1.04 $\pm$  0.1  & --     & --   & --    & --    & --    & --  & --  & --  \\
19     & 2500 & 3.749 $\pm$ 0.159  & 3.45 $\pm$  0.18 & --     & --   & --    & --    & --    & --  & --  & --  \\
20     & 2400 & 4.044 $\pm$ 0.181  & 0.42 $\pm$  0.42 & 0.008  & 100  & 100   & 1.137 & 1.125 & 21  & 20  & 45  \\
21     & 2200 & 2.814 $\pm$ 0.165  & 0.57 $\pm$  0.08 & --     & --   & --    & --    & --    & --  & --  & --  \\
22     & 2200 & 2.66  $\pm$ 0.165  & 0.39 $\pm$  0.13 & 0.4    & 3.16 & 1.0   & 1.0   & 1.0   & 11  & 12  & 30  \\
23     & 2700 & 18.74 $\pm$ 0.799  & 2.42 $\pm$  0.17 & --     & --   & --    & --    & --    & --  & --  & --  \\
24     & 2400 & 2.517 $\pm$ 0.139  & 2.14 $\pm$  0.16 & 0.4    & 31.6 & 31.6  & 1.05  & 1.05  & 20  & 18  & 60  \\
25     & 2000 & 2.8   $\pm$ 0.140  & 0.0  $\pm$  0.04 & 0.159  & 3.16 & 1.0   & 1.05  & 1.075 & 8   & 6   & 45  \\
26     & 2400 & 3.564 $\pm$ 0.171  & 0.4  $\pm$  0.1  & --     & --   & --    & --    & --    & --  & --  & --  \\
27     & 2300 & 2.862 $\pm$ 0.137  & 0.9  $\pm$  0.1  & --     & --   & --    & --    & --    & --  & --  & --  \\
28     & 3000 & 1.459 $\pm$ 0.076  & 2.89 $\pm$  0.16 & 0.008  & 3.16 & 0.32  & 1.0   & 1.025 & 11  & 12  & 30  \\
29     & 2200 & 20.46 $\pm$ 0.884  & 0.56 $\pm$  0.04 & 0.9    & 1.0  & 0.1   & 1.0   & 1.0   & 2   & 2   & 60  \\
30     & 2300 & 5.669 $\pm$ 0.237  & 1.43 $\pm$  0.07 & 0.013  & 3.16 & 3.162 & 1.05  & 1.025 & 11  & 8   & 45  \\[0.15cm]
\hline \myrule
31     & 2800 & 10.15 $\pm$ 0.612  & 0.95 $\pm$  0.32 & 0.044  & 0.32 & 0.32  & 1.20  & 1.20  & 18  & 20  & 45  \\
32     & 2800 & 47.78 $\pm$ 1.954  & 4.7  $\pm$  0.36 & 0.0377 & 10.0 & 10.0  & 1.05  & 1.075 & 8   & 10  & 45  \\
33     & 2800 & 82.84 $\pm$ 3.77   & 2.8  $\pm$  0.2  & 0.126  & 10.0 & 10.0  & 1.125 & 1.125 & 4   & 4   & 45  \\
34     & 2600 & 27.39 $\pm$ 0.995  & 15.0             & 0.051  & 316  & 316   & 1.175 & 1.175 & 20  & 20  & 45  \\
35     & 2800 & 51.18 $\pm$ 2.105  & 5.42 $\pm$  0.42 & 0.041  & 3.16 & 3.16  & 1.20  & 1.20  & 10  & 12  & 45  \\
36     & 1900 & 17.52 $\pm$ 0.933  & 2.31 $\pm$  0.33 & 0.051  & 10.0 & 3.16  & 1.025 & 1.05  & 12  & 12  & 45  \\
37     & 3000 & 64.0  $\pm$ 2.775  & 3.58 $\pm$  0.36 & 0.017  & 3.16 & 3.16  & 1.25  & 1.25  & 18  & 18  & 45  \\
38     & 2200 & 30.73 $\pm$ 6.521  & 7.74 $\pm$  0.48 & 0.025  & 0.32 & 1.0   & 1.10  & 1.125 & 18  & 18  & 60  \\
39     & 2400 & 26.27 $\pm$ 5.909  & 2.3  $\pm$  0.21 & 0.063  & 1.0  & 1.0   & 1.15  & 1.15  & 10  & 10  & 45  \\
40     & 3000 & 43.82 $\pm$ 1.729  & 4.3  $\pm$  0.29 & 0.015  & 0.1  & 0.1   & 1.05  & 1.025 & 4   & 4   & 30  \\
41     & 2800 & 23.81 $\pm$ 1.39   & 1.16 $\pm$  0.1  & 0.063  & 0.32 & 0.32  & 1.175 & 1.175 & 18  & 16  & 60  \\
42     & 2800 & 28.93 $\pm$ 5.848  & 0.04 $\pm$  0.01 & 0.025  & 10.0 & 10.0  & 1.075 & 1.075 & 18  & 14  & 60  \\
43     & 2100 & 12.48 $\pm$ 2.211  & 1.93 $\pm$  0.19 & 0.044  & 0.32 & 0.32  & 1.225 & 1.225 & 18  & 18  & 30  \\
44     & 2400 & 16.49 $\pm$ 5.241  & 1.02 $\pm$  0.26 & 0.018  & 1.0  & 1.0   & 1.225 & 1.225 & 20  & 18  & 30  \\
45     & 2800 & 34.59 $\pm$ 7.973  & 1.0  $\pm$  0.1  & 0.013  & 0.1  & 0.1   & 1.275 & 1.275 & 10  & 10  & 30  \\
46     & 3000 & 13.02 $\pm$ 3.103  & 0.6  $\pm$  0.1  & 0.021  & 0.32 & 0.32  & 1.175 & 1.175 & 12  & 12  & 60  \\
47     & 2500 & 6.242 $\pm$ 1.354  & 1.3  $\pm$  0.1  & 0.032  & 3.16 & 10.0  & 1.175 & 1.175 & 16  & 16  & 30  \\
48     & 2900 & 30.6  $\pm$ 7.081  & 0.3  $\pm$  0.1  & 0.012  & 1.0  & 1.0   & 1.15  & 1.15  & 14  & 14  & 45  \\
49     & 3400 & 27.66 $\pm$ 6.102  & 3.38 $\pm$  0.23 & 0.010  & 1.0  & 1.0   & 1.20  & 1.20  & 16  & 16  & 45  \\
50     & 2800 & 10.34 $\pm$ 2.422  & 1.3  $\pm$  0.1  & 0.007  & 31.6 & 31.6  & 1.125 & 1.15  & 18  & 18  & 30  \\
51     & 2800 & 14.88 $\pm$ 3.411  & 1.5  $\pm$  0.1  & 0.051  & 1.0  & 1.0   & 1.10  & 1.10  & 6   & 6   & 60  \\
52     & 2600 & 26.31 $\pm$ 6.167  & 0.6  $\pm$  0.1  & 0.01   & 0.32 & 0.32  & 1.20  & 1.225 & 10  & 10  & 45  \\
53     & 2600 & 6.486 $\pm$ 1.574  & 0.7  $\pm$  0.1  & 0.006  & 10.0 & 10.0  & 1.10  & 1.10  & 20  & 20  & 60  \\
54     & 3000 & 42.35 $\pm$ 9.502  & 1.7  $\pm$  0.1  & 0.014  & 0.3  & 0.32  & 1.20  & 1.20  & 8   & 8   & 30  \\
55     & 3000 & 21.45 $\pm$ 4.865  & 1.5  $\pm$  0.1  & 0.025  & 0.15 & 0.32  & 1.175 & 1.175 & 20  & 18  & 45  \\
\hline
\end{tabular}
\tablefoot{(1) We list the disk parameters for all objects in our sample with detected excess emission at IR wavelengths.
(2) The error in $T_{\rm{eff}}$ is 100\,K for all cases except for object 8 (i.e., OTS\,44) with a larger value 
of 140\,K. (3) We fixed $A_{V}=15\,\rm{mag}$ for object 34, see an explanation in Sect. \ref{sec:stellar}.
(4) The total disk mass $M_{\rm{disk}}$ is calculated from the dust mass assuming a gas-to-dust mass ratio of 100. 
(5) The tags $B$ and $P$ quoted in brackets for $M_{\rm{disk}}$, $\beta$ and $H_{100}$ denote the best-fit and the 
most probable values, respectively. The values of the disk inner radius $R_{\rm{in}}$ and the inclination $i$ are 
for the best-fit models.}
\end{center}
\end{table*}

\begin{figure*}[!ht]
  \centering
   \includegraphics[width=\textwidth,trim={0 20 0 0},clip]{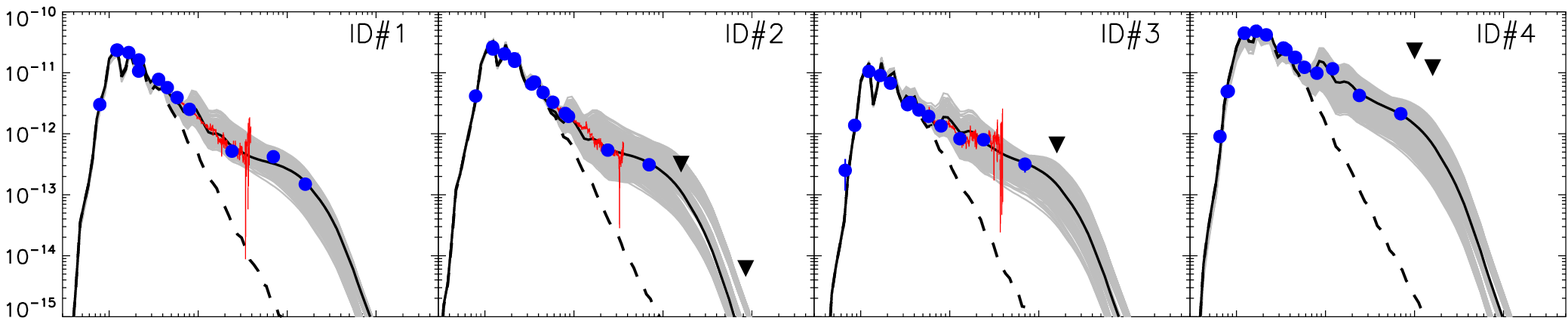} 
   \includegraphics[width=\textwidth,trim={0 20 0 0},clip]{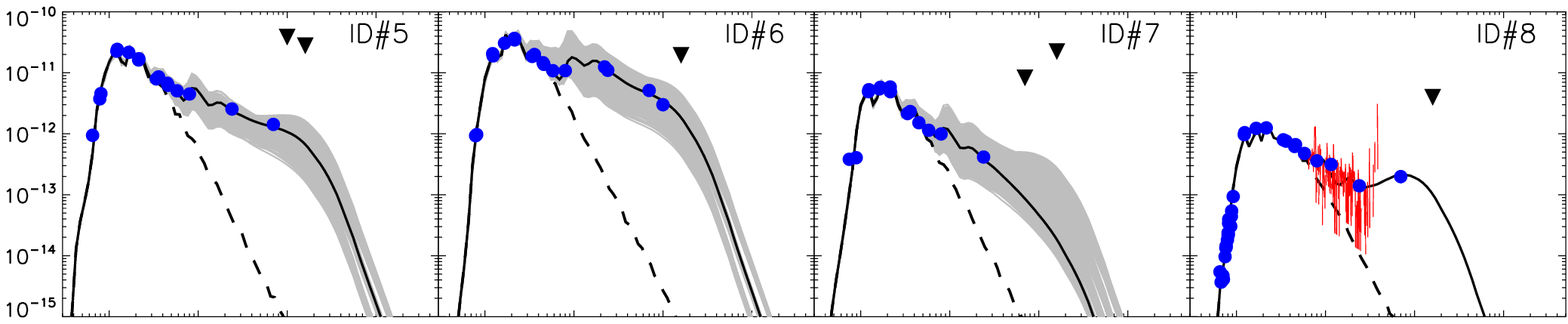} 
   \includegraphics[width=\textwidth,trim={0 20 0 0},clip]{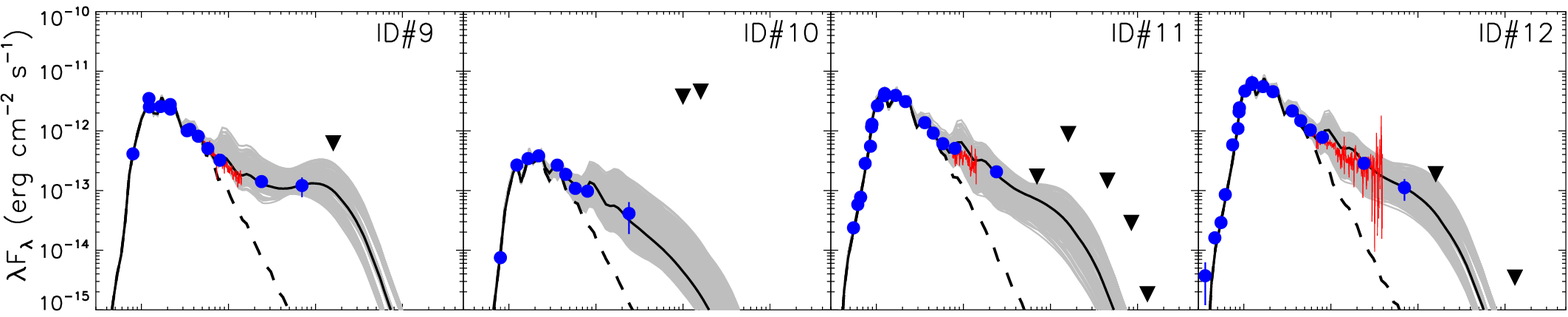}  
   \includegraphics[width=\textwidth,trim={0 20 0 0},clip]{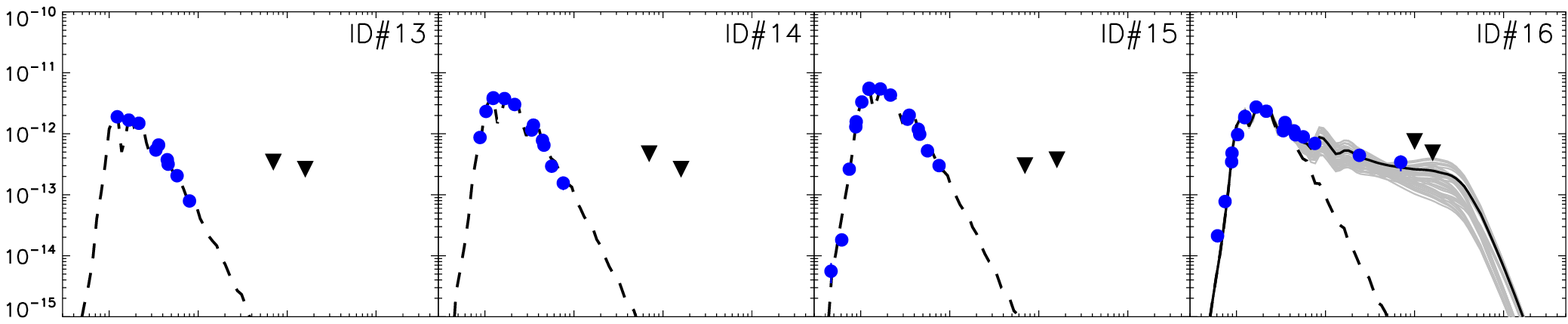}  
   \includegraphics[width=\textwidth]{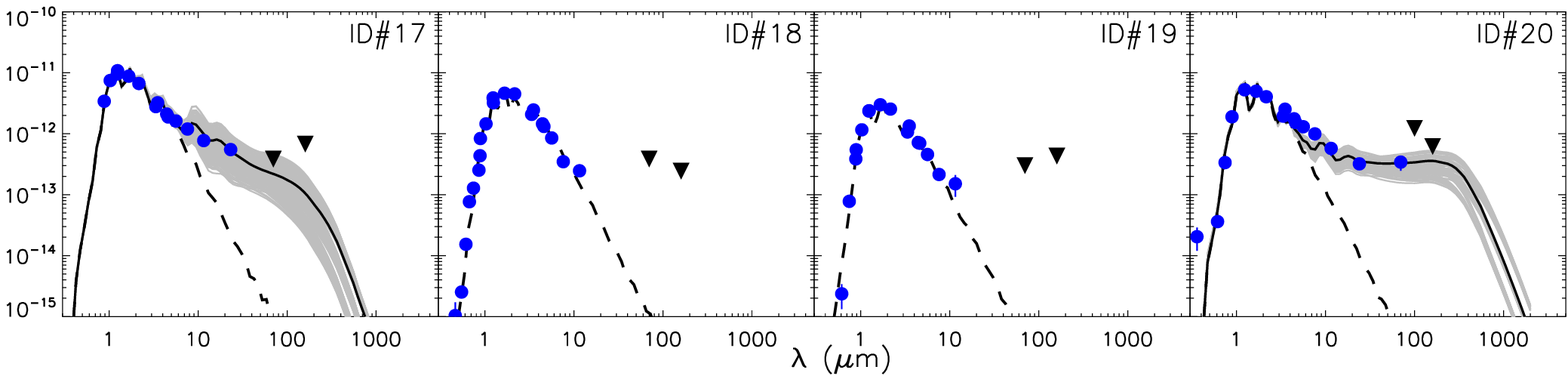}  
   \caption{Spectral energy distributions of the target disks. The dots depict the photometry at various wavelengths. 
            {\it Spitzer}/IRS spectra are indicated as red lines. The upside down triangles show the $3\,\sigma$ upper limits of 
            the flux density. The best-fit models are indicated as black solid lines, whereas the dashed lines represent the 
            photospheric emission levels. The gray lines denote all probable model fits that are derived by Bayesian 
            analysis, showing the strength of constraints of the best fit models for each object, see Sect. \ref{sec:fit_method}.}
\label{fig:bestfit}
\end{figure*}

\addtocounter{figure}{-1}

\begin{figure*}[!ht]
  \centering
   \includegraphics[width=\textwidth,trim={0 20 0 0},clip]{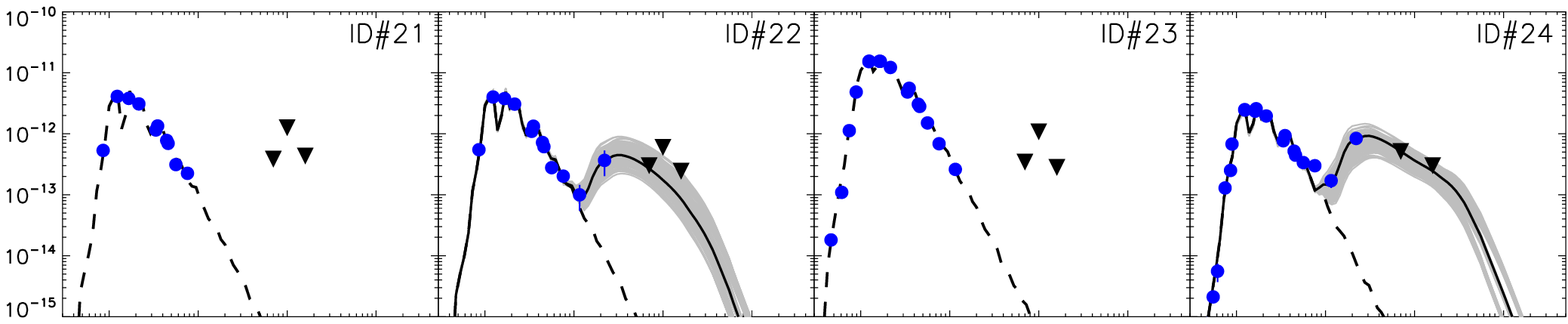}  
   \includegraphics[width=\textwidth,trim={0 20 0 0},clip]{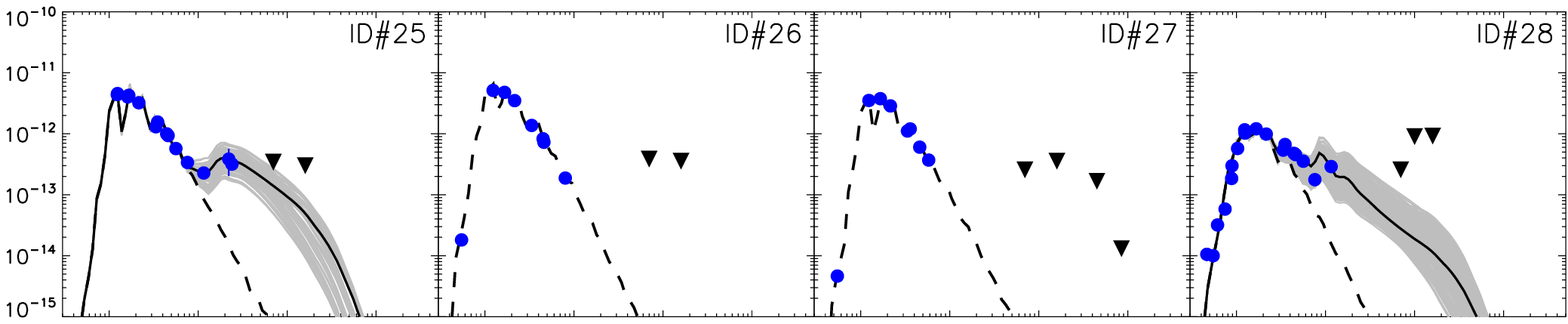}  
   \includegraphics[width=\textwidth,trim={0 20 0 0},clip]{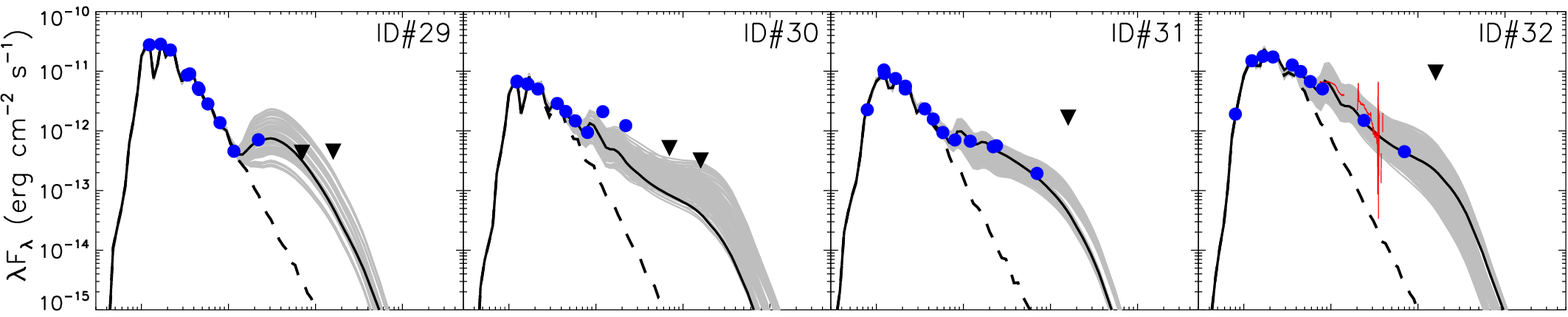} 
   \includegraphics[width=\textwidth,trim={0 20 0 0},clip]{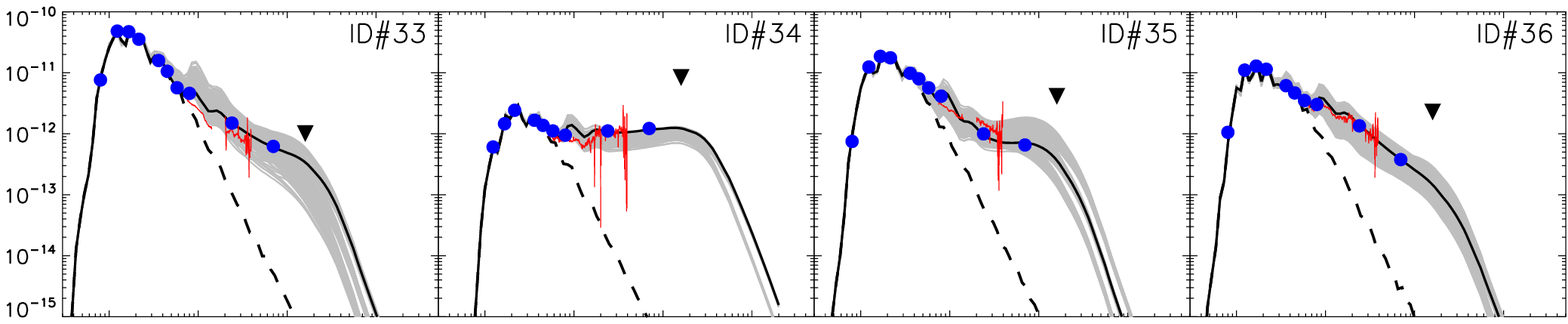}
   \includegraphics[width=\textwidth]{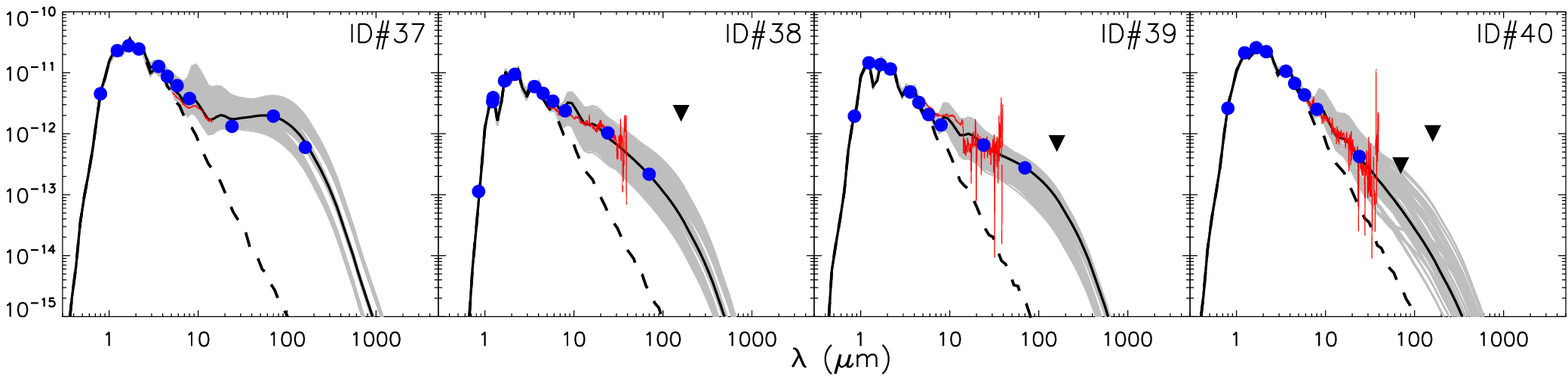}
   \caption{Continued.}
\end{figure*}

\addtocounter{figure}{-1}

\begin{figure*}[!ht]
  \centering
   \includegraphics[width=\textwidth,trim={0 20 0 0},clip]{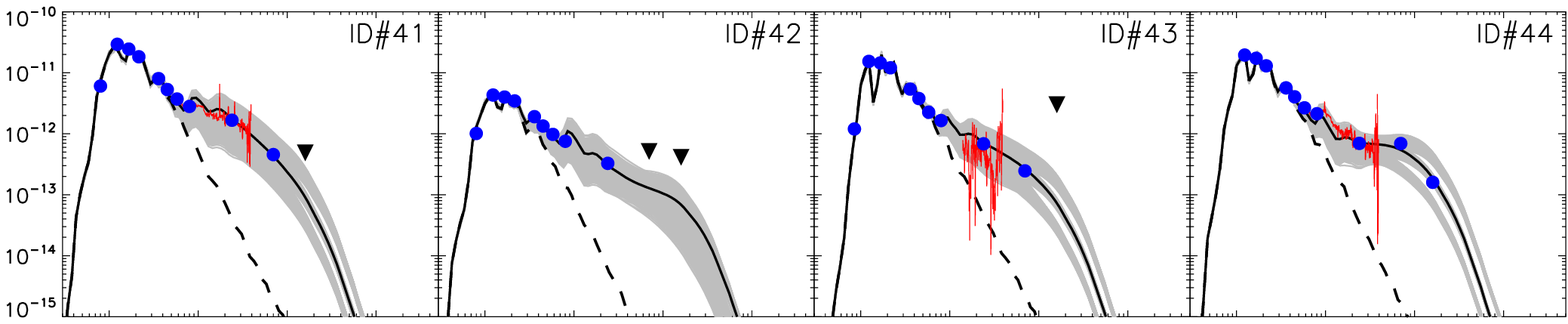}
   \includegraphics[width=\textwidth,trim={0 20 0 0},clip]{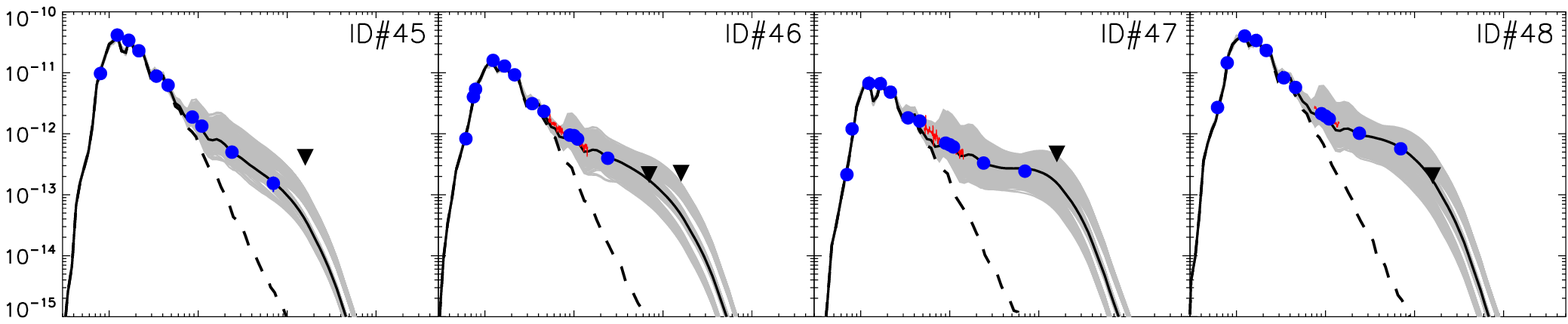}
   \includegraphics[width=\textwidth,trim={0 20 0 0},clip]{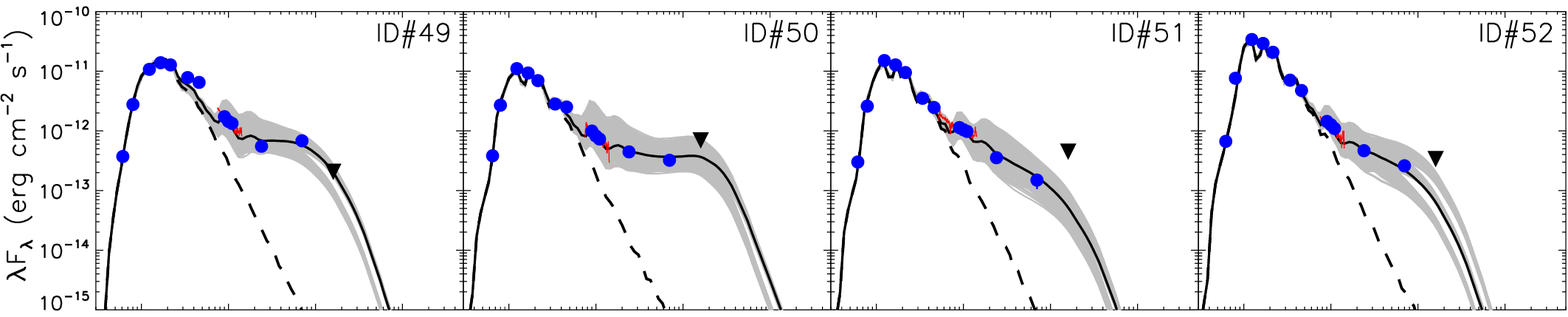}
   \includegraphics[width=\textwidth]{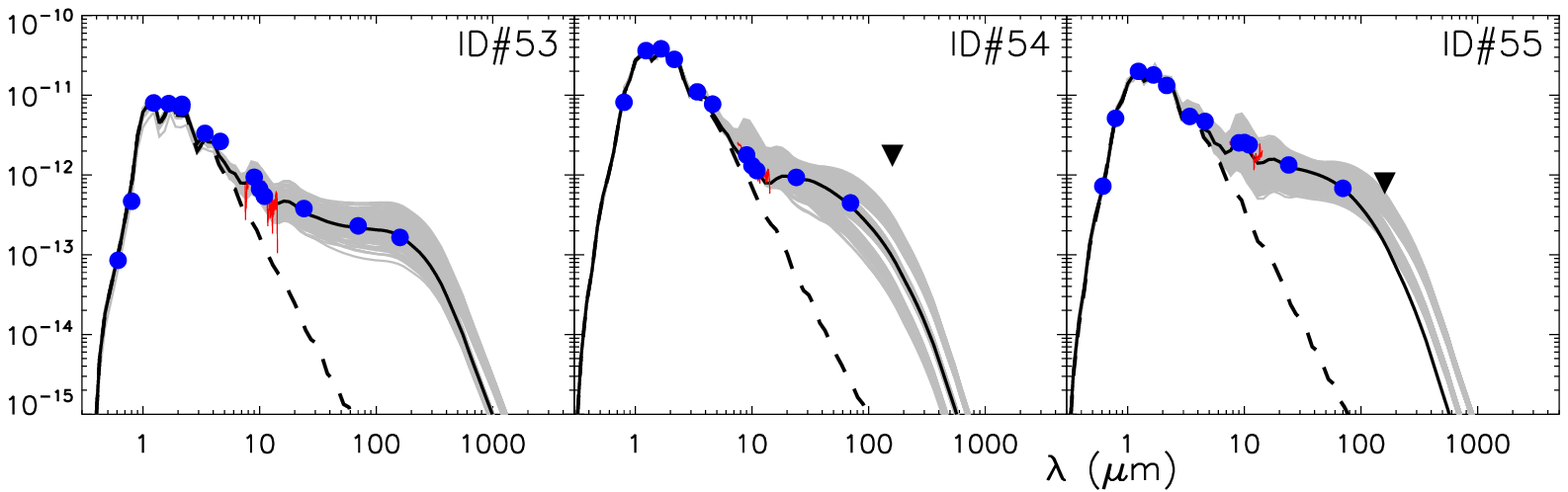}
   \caption{Continued.}
\end{figure*}

\subsection{Fitting method}
\label{sec:fit_method}
The task of SED fitting was conducted with a grid of pre-calculated models for each target. 
The model grids can give us an overview of the fitting quality in different parameter domains and 
clues to improve the quality of the fit with further attempts if necessary. Table \ref{tab:parasam} summarizes 
the explored ranges of each parameter considered here. The grid adds up to a total number of 37\,440 models for each source.
We tried to achieve a better explanation on the observations on one hand by reducing the dimensionality 
of the parameter space as far as possible and on the other hand by using a denser grid of models as compared
to previous studies on {\it Herschel}/PACS data of BD disks \citep[e.g.,][]{harvey2012a,harvey2012b,alves2013,olofsson2013,spezzi2013}.
For objects, for which we could not obtain good results from the grid, simulated annealing approach was additionally invoked 
to improve the solution by taking the best model in the grid as the starting point of the Markov chain.
Generally, ${\sim}\,1000$ steps for the local fit with simulated annealing are enough \citep{lium2013}. This kind of methodology 
makes use of the advantages of both database method and simulated annealing algorithm and has already been demonstrated to be 
successful for SED analysis \citep{lium2012, madlener2012}.

\begin{table}[!t]
 \centering
    \caption{Parameter range of the model grids computed in this work.}
    \begin{tabular}{lccccc}
     \hline
     \hline
   Parameter                        &  Min &  Max  &  Number of values  &  Sampling  \\
     \hline
   $R_{\rm in}$ [$r_{\rm{sub}}$]    & 1    & 100     &  6 & logarithmic   \\ 
   $\beta$                          & 1.0  & 1.3     & 13 & linear   \\
   $H_{\rm 100}$ [AU]               & 2    & 20      & 10 & linear  \\
   log\,$M_{\rm{disk}}$ [$\rm M_{\odot}$]  & $-$6 & $-$2.5 &  8 & logarithmic   \\
   $i$ [$^{\circ}$]                 & 15   & 90      &  6 & linear  \\
   \hline \\
   \end{tabular}
\label{tab:parasam}
\end{table}

The fitting results are displayed in Figure \ref{fig:bestfit}. The best-fit models are indicated as black solid lines, whereas 
the dashed lines represent the photospheric emission level. The corresponding disk parameter sets are listed in Table \ref{tab:param}. 
The {\it Spitzer}/IRS spectra, when available, are taken into account by fitting the subjacent continuum. Since our goal is to 
characterize the overall structure of the disks, we therefore did not attempt to reproduce the exact shape of the silicate 
feature that is mainly related to silicate mineralogy of the dust in the disk atmosphere \citep[e.g.,][]{bouwman2008, olofsson2010}. 
We emphasize that the best fits presented here cannot be considered as a unique solution due to the degeneracy of SED models. 
Despite this, previous studies have shown that modeling SEDs with broad wavelength coverage can constrain the mass and 
geometry of disks around BDs \citep[e.g.,][]{harvey2012a, harvey2012b, olofsson2013, spezzi2013}. 
In particular, the Bayesian inference can analyze the potential correlations and interplay between different 
parameters in a statistical way \citep[e.g.,][]{pinte2008}. We derived the Bayesian probability distributions 
of the disk parameters, from which we calculated the most probable values and estimated the validity ranges for 
each parameter, corresponding to regions where $P>0.5 \times P_{\rm Max}$ \citep[e.g.,][]{liu2015}. 
The gray lines in Figure \ref{fig:bestfit} denote all the models that are within the validity ranges.

\section{Discussion}
\label{sec:discussion}
We re-analyzed the {\it Herschel}/PACS data of a sample of 55 BDs and very low mass stars in a homogeneous way and constructed 
their observed SEDs by complementing our {\it Herschel} photometry with previous flux measurements available at wavelengths ranging 
from optical to (sub-)millimeter. For the 46 targets that show IR excess emission, we characterized the properties of their 
surrounding disks using radiative transfer technique under identical assumptions for the modeling setup. 
Moreover, we evaluated the constraints on different disk parameters through Bayesian analysis. 

Despite the good wavelength coverage of observations and efforts to reduce the degree of freedom, the degeneracy of SED models
still exists. Therefore, we statistically analyzed the most probable results from Bayesian statistics, other than the best-fit 
models, because the former case is likely more representative of the overall fitting quality. The most probable parameter sets 
for each target are identified as the peak of Bayesian probability distribution. The derived values are summarized in 
Table \ref{tab:param}.

\subsection{Overview of modeling results}

We characterized the studied disks by determining their inner radii $R_{\rm{in}}$, flaring indices $\beta$, 
scale heights $H_{100}$, disk masses $M_{\rm{disk}}$ and inclinations $i$. The results are listed in Table \ref{tab:param}.
We give an overview of the statistical behaviour of these properties in the following. 

The disk inner radius can be well constrained for all objects because the observed SEDs are well 
sampled in the near- and mid-IR domains. The inner radii of the disks are likely close to the dust sublimation
radii $r_{\rm{sub}}$, which are of the order of [0.005$-$0.05\,AU] from the faintest to the brightest targets in our sample.
We found four candidates of BD transition disks with $R_{\rm{in}}>20\,r_{\rm{sub}}$, i.e., source 22 (J04302365+2359129), 
source 24 (J04355143+2249119), source 25 (J04361030+2159364) and source 29 (J04221332+1934392). 
Current data indicate no detectable excess shortward of $12\,\mu{\rm{m}}$ accompanying with a steep slope 
in the mid-IR of these four objects. However, none of them is detected at {\it Herschel}/PACS bands. 
Follow-up observations, in particular spatially resolved millimeter images, are required to clarify their
evolutionary stages \citep[e.g.,][]{williams2011,andrews2011}. The inclination of the disk cannot be 
constrained very well with the available SED data as shown by the flat probability distribution of this 
parameter nearly for all targets. Accurate determination of disk inclinations requires spatially resolved 
observations \citep[e.g.,][]{chiang1999, robitaille2007}.   

The geometric structure of the disk is described by the flaring index $\beta$ and scale height $H_{100}$ according 
to the model ansatz, see Sect. \ref{sec:model}. Our detailed SED analysis places tight constraints on the disk geometry, 
because the best-fit $\beta$ and $H_{100}$ are in good agreement with the most probable values (see Table \ref{tab:param}) 
and the probability distribution functions are clearly non-flat for most objects (see the Appendix). We observed a 
preferential values of $\beta$ in the range $1.05-1.20$, while the majority of the target disks are best modeled with 
$H_{100}$ of the order of $5-20\,\rm{AU}$. As shown by \citet{harvey2012a} and \citet{liu2015}, most parts (except for 
the inner region) of typical BD disks produce optically thin emission at far-IR wavelengths, demonstrating the applicability 
of PACS photometry as a diagnosis for the disk mass of BDs. Objects with detections at 70 and/or 160μm allow reasonable 
determination of their disk masses as demonstrated by clear peaks in the Bayesian probability distributions of $M_{\rm{disk}}$ 
in these cases. For the remaining targets, the disk mass is not well constrained, typically showing two or three 
indistinguishable peaks in the probability distributions. Overall, the likely disk masses feature a wide range 
of [$10^{-6}, 10^{-4}\,M_{\odot}$] with a median value of the order of $10^{-5}\,M_{\odot}$, where we assume the dust 
properties as described in Sect. \ref{sec:model} and a gas-to-dust mass ratio of 100. Our statistical results on the disk 
parameters of substellar hosts are widely in agreement with previous findings from case studies and surveys, however, for 
individual sources presented here our measured PACS flux densities and correspondingly the model solutions differ from those of 
previous works \citep[e.g.,][]{harvey2012a, harvey2012b, riaz2012a, alves2013, joergens2012, joergens2013, spezzi2013}. 
A few BDs detected at (sub)millimeter to date and their model explanations also support our results on the typical 
structural properties especially the mass of BD disks \citep[e.g.,][]{klein2003,scholz2006,mohanty2013,ricci2012,ricci2013,ricci2014,broekhoven2014}.
 
\subsection{Comparison with disks around young stellar objects}
\label{sec:comparison}

Comparing the derived disk properties with those of their higher mass counterparts, such as T Tauri disks,
can provide important insights into understanding the formation of BDs and very low mass stars. 
Several attempts have been made in this direction \citep[e.g.,][]{pascucci2009, szucs2010}. 
Plenty of multi-wavelength data including spatially resolved ones exist for young stellar objects, 
leading to robust constraints on the structure and mass of their disks.  
In particular, coherent multi-wavelength modeling suggests a typical flaring index $\beta \sim 1.25$ and a favorable 
range $5-20\,\rm{AU}$ of $H_{100}$ for T Tauri disks \citep[e.g.,][]{wolfp2003,walker2004,sauter2009,madlener2012,grafe2013,garufi2014}. 
Based on (sub-)millimeter measurements of a large sample of young stellar objects in various star formation regions, 
the disk masses of T Tauri stars are found in the range of [$10^{-3}, 10^{-1}\,M_{\odot}$] \citep[e.g,][]{andrews2007,lee2011,sicilia2011, andrewsr2013, carpenter2014}.    

We found significant evidence for different properties of disks between sun-like and cool stars, showing that disks around
BDs and very low mass stars are generally flatter and orders of magnitude less massive than T Tauri disks. Although similar 
conclusions were drawn by previous studies \citep{pascucci2009, szucs2010, olofsson2013}, the evidence we found should be 
more significant because our results are based for the first time on a homogenous analysis all the way from {\it Herschel} 
data reduction and flux density measurement to SED modeling, and to interpretation. Theoretical models predict that disks 
around cooler stars should be more extended in the vertical direction (Walker et al. 2004). However, our results, together 
with previous studies \citep[e.g.,][]{harvey2012a, alves2013, olofsson2013}, indicate that the disk scale height is independent 
of the host stellar mass. Dust growth and settling are the two pivotal physical processes simultaneously shaping the 
disk structure, for instance reducing the disk scale height \citep{dalessio2006}. Mid-IR observations have shown that these 
processes probably occur in early stages of the disk evolution \citep[e.g.,][]{furlan2006,mcclure2010}. Given the broad age spread, 
our targets may be at different evolutionary stages during which the degrees of dust growth and settling are not at the same level. 
This will probably weaken any relation, if present, between the scale height and the (sub-)stellar mass.  
Nevertheless, analyses of coeval samples, for example in Cha\,I \citep{olofsson2013} and $\rho$ Oph \citep{alves2013},
also indicate that disk scale heights are comparable in both sun-like stars and BDs. Future high-resolution (sub-)millimeter 
observations are indispensable to quantify the degree of dust settling and better constrain the disk scale 
height \citep[e.g.,][]{sauter2009,boehler2013}.

\subsection{Dependence of disk parameters on the mass of substellar hosts}

There is growing observational evidence that disk evolution depends on the mass of the central object. For example, the primordial 
dust disks are believed to live longer for cool stars, because the observed disk frequency of cool stars and BDs is 
higher in comparison with sun-like stars \citep[e.g.,][]{carpenter2006,scholz2007,riaz2008,bayo2012,ribas2015}.  
The observed trend of small disk accretion rates of BDs indicates that the scaling between the accretion rate and 
stellar mass (i.e., $\dot{M}_{\rm{acc}} \propto M_{\star}^{{\sim}2}$) deduced mainly from low- and intermediate-mass 
stars extends to the substellar mass spectrum \citep[e.g.,][]{muzerolle2005,mohanty2005,herczeg2009,bayo2012}. 
Analysis of {\it Spitzer}/IRS spectra demonstrates that the dust processing, such as grain growth and crystallization, in 
the mid-IR emitting regions of BDs and very low mass stars appears to be more advanced than in disks around sun-like 
stars with similar age \citep[e.g.,][]{apai2005,pascucci2009,riaz2009}. Moreover, \citet{pascucci2013} also found 
stellar-mass-dependencies in the atomic and molecular content of disk atmospheres.

Searching for evidence of stellar-mass-dependent disk properties is very common and interesting, 
because it has important implications for planet formation theories. \citet{szucs2010} divided ${\sim}200$ young stellar 
objects and BDs in the Chamaeleon I star-forming region into two groups with SpTs earlier or later than
$\sim$M4.5. Through modeling the median SEDs (longward of MIPS\,$24\,\mu{\rm{m}}$) of these two groups, they found 
that the disk is on average flatter in the lower stellar mass group than in the group with higher stellar masses. 
Based on {\it Spitzer} data, several other works hinted at similar results \citep[e.g.,][]{riaz2012b}. 
As laid out in Sect. \ref{sec:comparison}, our results derived by including the {\it Herschel} far-IR observations 
evince the same tendency. Note that this is a statistical behaviour shown from the 
ensemble of the 46 modeled very low-mass stars and BDs.

\begin{figure}[!t]
\includegraphics[width=0.45\textwidth]{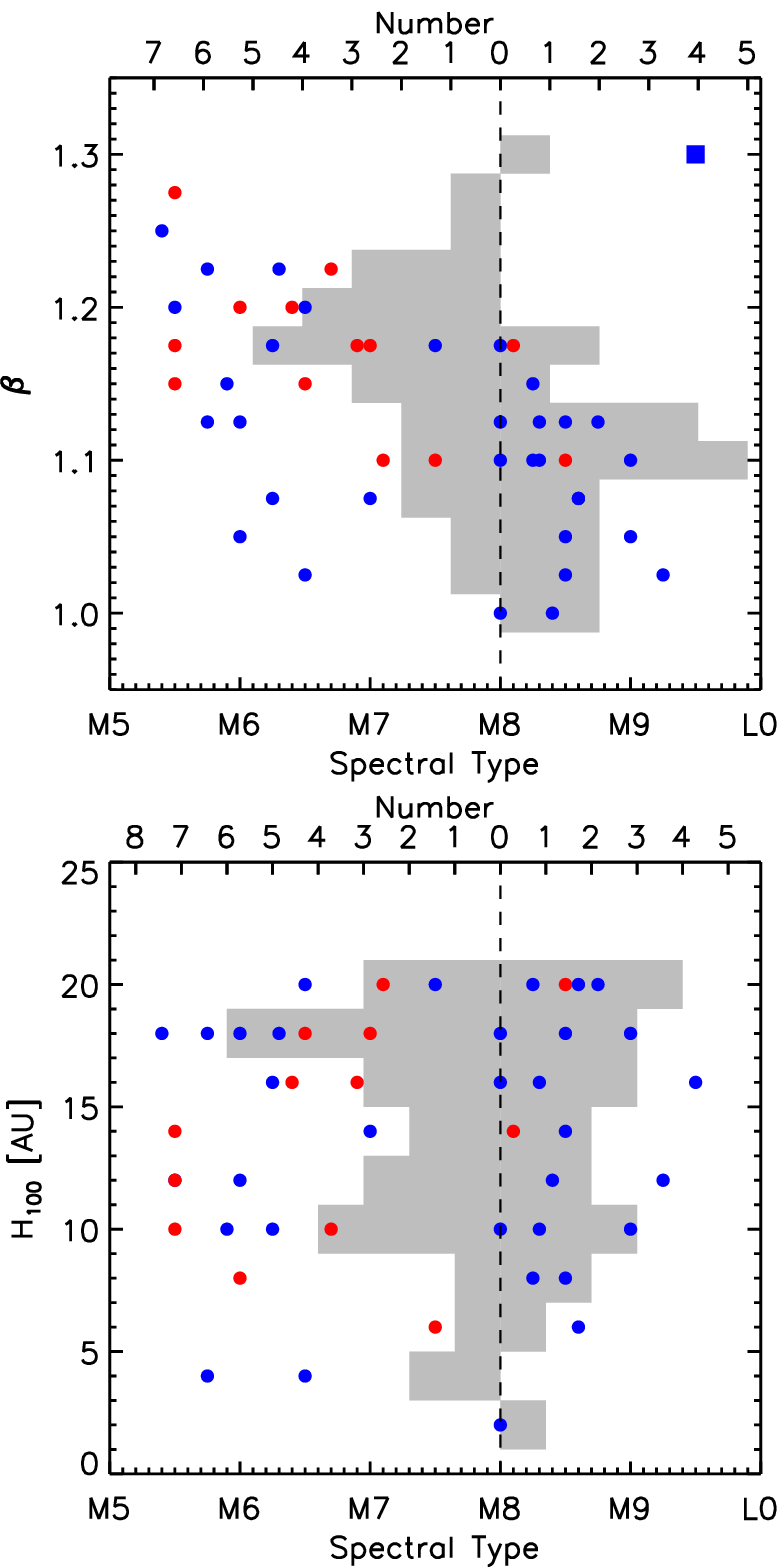}
   \caption{Flaring index $\beta$ (top) and scale height $H_{100}$ (bottom) as a function of SpT. 
            We found that $\beta$ decreases clearly from early-type BDs to late-type BDs (top), while 
            $H_{100}$ appears independent of the spectral type in the BD regime. The inserted diagrams in 
            each panel show the histograms of these two parameters calculated for targets that belong to the ETBD and 
            LTBD groups respectively. The vertical dashed lines symbolize the criterion (i.e., M8) used to 
            divide the targets, see Sect. \ref{sec:sample}. Blue dots represent objects located in clusters with
            ages ${<}5\,\rm{Myr}$, whereas the red ones refer to older (${\gtrsim}5\,\rm{Myr}$) targets. 
            The blue square in the upper panel points to the obvious outlier target (OTS\,44) of the observed 
            relation. For better representation, slight changes in either the SpT or disk parameters are made for 
            objects with a pair of identical values.}
\label{fig:spt_evo}
\end{figure}

As a further step towards examining whether the trends of disk properties extend to the low-end of the 
substellar mass regime, we performed a detailed comparison of the fitted disk geometry parameters between the ETBD and 
LTBD groups as introduced in Sect. \ref{sec:sample}. The results are displayed in Figure \ref{fig:spt_evo}, in which the upper panel shows a 
clear decrease of disk flaring from early to late type BDs, with a typical dispersion of the order 
of ${\sim}0.1$. The histogram inside the figure confirms the relationship between the flaring index $\beta$ and 
SpT. In particular, the median value of $\beta$ is 1.10 in the LTBD group, while the ETBD 
group features a higher median value of 1.175. The mean value of $\beta$ is also found to be smaller in the 
LTBD group, i.e., $\beta_{\rm{mean}}=1.10$, as compared to $\beta_{\rm{mean}}=1.16$ for the ETBD group. 
The difference in the median/mean disk flaring between the two groups is $\Delta\,\beta \sim 0.07$, which is about 
three times of the step size for this dimension used to build the model grid. 
The lower $\beta$ of cooler BDs suggests that disk flaring is very sensitive to far-IR photometry, because the LTBD
group is generally fainter at $70\,\mu\rm{m}$, see Sect. \ref{sec:obs_trend}. 
OTS\,44, an M9.5 BD with an estimated mass of ${\sim}12\,M_{\rm{Jup}}$ in the Cha\,I star-forming region
\citep{bonnefoy2014}, is an obvious outlier of the observed relation, see the blue square in the upper panel 
of Figure \ref{fig:spt_evo}. The significant IR excess together with a rising SED 
from 24 to $70\,\mu\rm{m}$ require a highly flared disk to reproduce the data \citep[][this work]{joergens2013, joergens2015}.
We note that our re-measurement of the {\it Herschel}/PACS flux densities of OTS\,44 leads to a slightly smaller disk mass of 
$3.2\times10^{-5}\,M_{\odot}$ \citep[${\sim}10\,M_{\oplus}$;][this work]{joergens2015} compared to a previous estimate \citep[${\sim}30\,M_{\oplus}$;][]{joergens2013} 
based on the photometry given by \citet{harvey2012a}. VLT/SINFONI spectra revealed active mass accretion for this extremely 
low-mass object \citep{joergens2013}, suggesting that OTS\,44's disk is probably at its early stage of evolution, during which 
dust settling (a key role of reducing $\beta$) has not yet proceeded very far. The observed $\rm{SpT}-\beta$ dependency 
is unlikely a bias of the mixed ages and evolutionary stages of the sample, since both the young ($<5\,\rm{Myr}$, blue points) 
and old ($\gtrsim5\,\rm{Myr}$, red points) subgroups display the same tendency, see Figure \ref{fig:spt_evo}.

As shown in the lower panel of Figure \ref{fig:spt_evo}, there is no clear 
trend visible in the distribution of scale height. The median or mean values of $H_{100}$ are close to each other 
for both the ETBD and LTBD groups, i.e., $H_{100}{\sim}14\,\rm{AU}$. This means that the independency of $H_{100}$ 
on SpT is also found in the low-mass substellar regime. The disk scale height varies like $H\,{\propto}\,(\langle T_{d}\rangle/M_{\star})^{0.5}$ 
if the hydrostatic equilibrium between the dust and gas phases is assumed, where $\langle T_{d}\rangle$ is the density weighted mean dust temperature. 
\citet{andrewsr2013} found that $\langle T_{d}\rangle$ scales with the host stellar luminosity basically according to $\langle T_{d}\rangle\,{\approx}\,25 (L_{\star}/L_{\odot})^{1/4}\,\rm{K}$. 
The stellar luminosities $L_{\star}\,{\propto}\,M_{\star}^{1.5-2}$ in their sample that covers the substellar regime. 
This suggests that the correlation between the scale height and $M_{\star}$ may be intrinsically not tight. 
The scaling between the scale height and $M_{\star}$ highly depends on the detailed coupling (i.e., dust settling) 
between the gas and small dust grains, which is quite complicated and has to be investigated with future 
high-resolution (sub-)millimeter observations \citep[e.g.,][]{sauter2009, boehler2013}.

\begin{figure}[!t]
\includegraphics[width=0.45\textwidth]{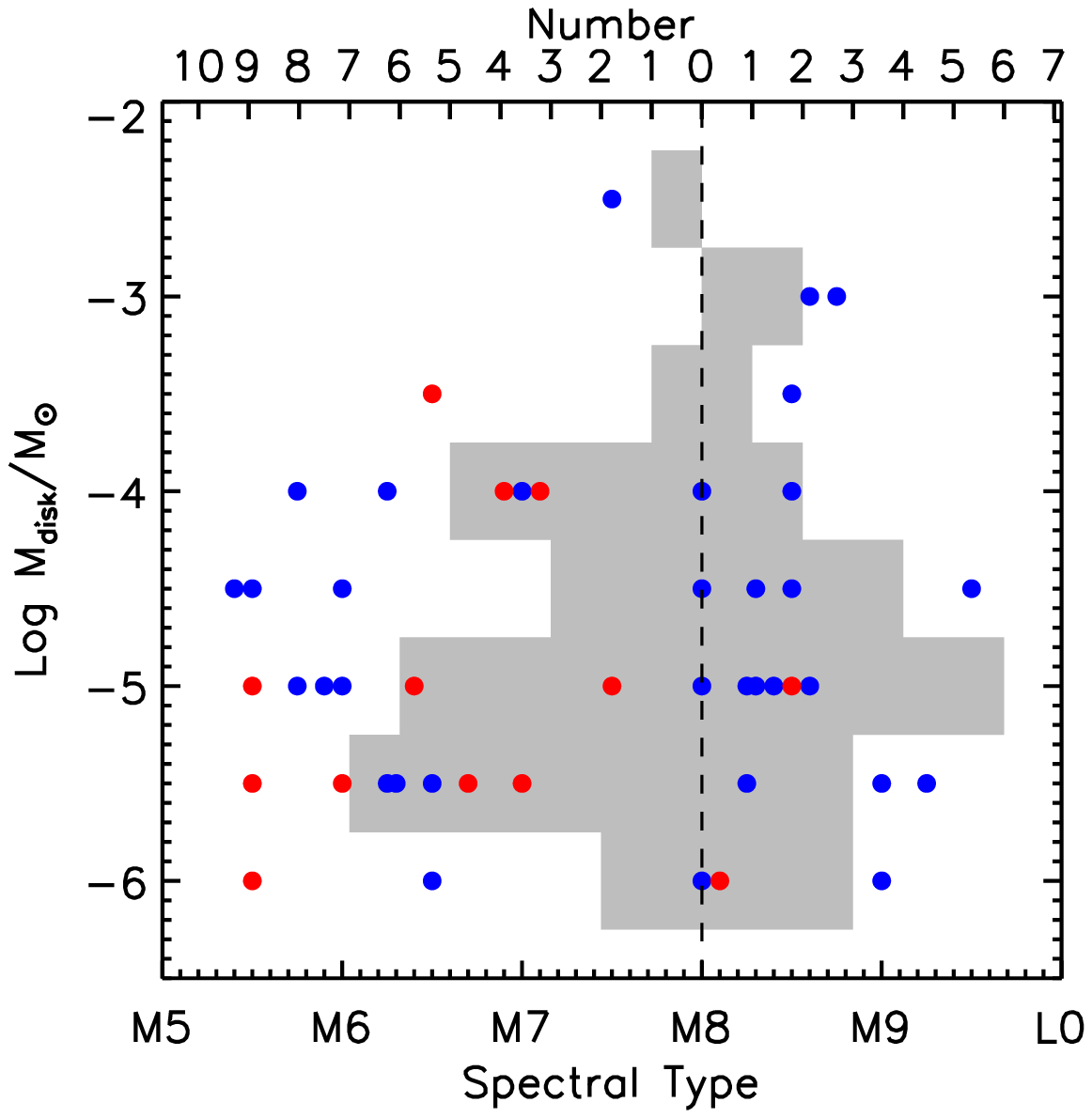}
   \caption{Same as in Figure \ref{fig:spt_evo}, but for the disk mass $M_{\rm{disk}}$.}
\label{fig:spt_mdisk}
\end{figure}

The median disk masses for both the ETBD and LTBD groups are $M_{\rm{disk}} = 10^{-5}\,M_{\odot}$, a value 
essentially consistent with the one given by \citet{harvey2012a}. As shown in Figure \ref{fig:spt_mdisk}, we did not 
find an obvious correlation between the disk mass and SpT (as a proxy for the mass of the central object) in the BD 
mass regime, which seems to deviate from the speculation from the tight correlation between the PACS\,70\,$\mu{\rm{m}}$ 
flux density and SpT shown by \citet{liu2015}. This is probably due to one or more of the following reasons:
(1) the constraints on the disk mass are mostly from the 70\,$\mu{\rm{m}}$ photometry in our study.
    However, only $52\%$ (i.e., 11/21) of the modeled LTBD objects are detected at $70\,\mu{\rm{m}}$, 
    whereas the detection rate is $88\%$ (i.e., 22/25) for the ETBD group, see Table \ref{tab:sample}. 
    The (most probable) disk masses of the undetected sources are quite uncertain, potentially preventing 
    the underlying correlation, if present, from showing up. 
(2) The conversion from PACS\,70\,$\mu{\rm{m}}$ flux density to disk mass is not straightforward. On one hand 
    the target disks are not completely optically thin at PACS wavelengths \citep{harvey2012a}, and on the 
    other hand 70\,$\mu{\rm{m}}$ is not in the clear Rayleigh-Jeans tail of the SED.
(3) Our correlation analysis is based on a sample of targets that span a wide range of ages. 
    Therefore, the targets are probably at different evolutionary stages, which may weaken any relation between the
    (sub)stellar and disk masses. Beside the issue of incompleteness of our sample selection in each cloud, the detection 
    bias can also affect the correlation because more distant clouds have a worse sensitivity to disk mass compared 
    to closer siblings.       

\begin{figure}[!t]
\includegraphics[width=0.5\textwidth]{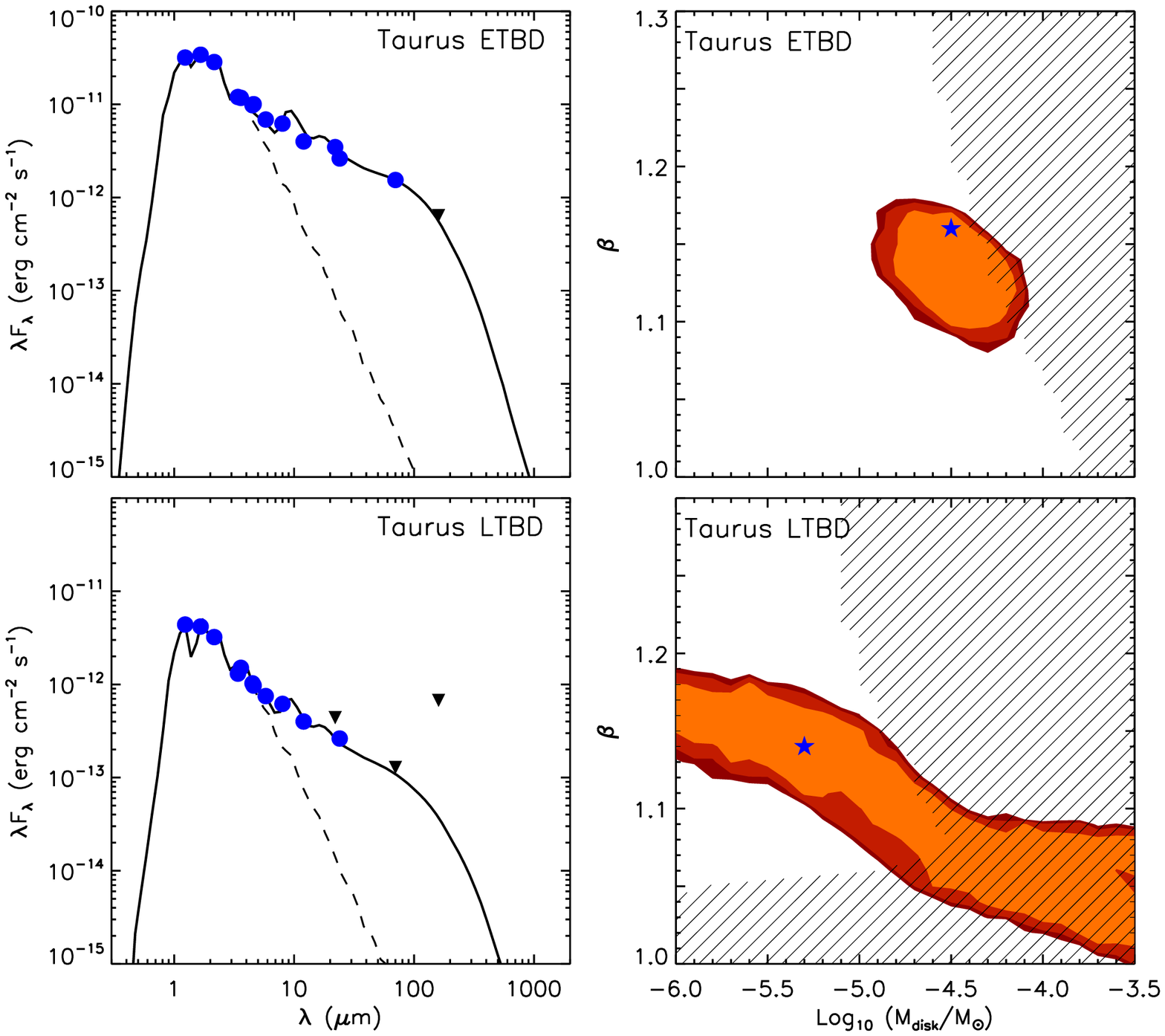}
   \caption{{\it Left panels: } Median SEDs of Class II ETBDs (upper) and LTBDs (lower) in the Taurus molecular cloud.   
            The dots depict the median flux densities at various bands. The upside down triangles show that the median 
            flux densities at the corresponding wavelengths are treated as upper limits. The best fit models are indicated as solid
            lines, whereas the dashed lines mark the photospheric emission levels. In the fitting procedure, the disk scale 
            height $H_{100}$ and inclination $i$ are fixed to $14\,\rm{AU}$ and $45^{\circ}$ respectively. 
            The parameter set of the best model in the upper panel: 
            $R_{\rm{in}}=0.09\,\rm{AU}$, $M_{\rm{disk}}=3\times10^{-5}\,M_{\odot}$, $\beta=1.16$. The parameter values of 
            the best model in the lower panel: $R_{\rm{in}}=0.03\,\rm{AU}$, $M_{\rm{disk}}=5\times10^{-6}\,M_{\odot}$, $\beta=1.14$.
             {\it Right panels:} 2D contour plots of 
            the $\chi^2$ function projected over disk mass $M_{\rm{disk}}$ and flaring index $\beta$. Contours with different 
            shades of red are drawn at the 68\%, 90\%, 95\% 2D confidence intervals. The blue stars represent the best-fit disk 
            models. The hashed regions correspond to models that overpredict the upper limits of the median flux density.}
\label{fig:tau_median}
\end{figure}

In order to further investigate the $M_{\rm{disk}}-\rm{SpT}$ relation in the substellar mass regime, we compile 
the median SEDs of ${\rm Class\,\,II}$ Taurus sources with SpTs lying in the same range defined in our sample 
selection. Our compilation is entirely based on the catalog as reported by \citet{bulger2014}. We only consider sources 
that have flux density measurements including upper limits available at all the 2MASS J/H/K bands, four IRAC and WISE bands, 
MIPS\,$24\,\mu{\rm{m}}$, and PACS\,70 and $160\,\mu{\rm{m}}$. With this criterion, there are in total 32 targets entering 
in the analysis. 25 sources are ETBDs, whereas 7 targets are LTBDs. The Kaplan-Meier product-limit estimator 
is used to properly account for upper limits, if encountered, in the estimation of median flux density. All the PACS flux 
densities are taken from \citet{bulger2014} for homogeneity, although we re-performed photometry for some Taurus targets. 
The median SEDs of Taurus ETBDs and LTBDs are shown as blue dots in Figure \ref{fig:tau_median}. An important result is that the
median $F_{70}=36\,\rm{mJy}$ of Taurus ETBD group is at least twelve times larger than that of the Taurus 
LTBDs, i.e., $F_{70}<3\,\rm{mJy}$. The difference in $70\,\mu{\rm{m}}$ flux density between early- and late-type targets becomes 
significantly larger in coeval planet-forming disks as compared to the result from our sample of objects with different ages,
see Sect. \ref{sec:obs_trend}. We fit the median SEDs by using the disk and dust models as described in Sect. \ref{sec:model}.
As we discussed above, the disk scale height $H_{100}$ is independent on SpT. Hence, we fixed $H_{100}$ to 
14\,AU, i.e., the aforementioned median value of $H_{100}$ of our sample. The disk inclination $i$ was set to $45^{\circ}$ 
because SED modeling cannot provide tight constraints on this parameter. These assumptions were made in order to further 
reduce the model degeneracy, and therefore to focus on the parameter study between $\beta$, $M_{\rm{disk}}$ and SpT.
Figure \ref{fig:tau_median} shows the fitting results and 2D contour plots of the $\chi^2$ function projected over $M_{\rm{disk}}$ 
and $\beta$. The flaring index $\beta$ is not well constrained in Taurus LTBD group, with probable values 
extending to a lower level than the case for Taurus ETBDs. This may be an indicator that disks around lower mass 
objects are indeed less flared in general. The best fit value together with the confidence intervals clearly indicate that 
Taurus ETBDs are very likely more massive than Taurus LTBDs, which appears discrepant with the theoretical predictions 
of disk fragmentation models \citep{stamatellos2015}. However, further efforts are highly desired to clarify this issue
given the fact that the disk mass presented here is roughly estimated with far-IR data. With its unprecedented 
performance, ALMA can observe very faint disks at longer wavelength and provide robust measurement of disk 
mass \citep[e.g.,][]{ricci2014}, so as to testify whether the BD disks obey the same $M_{\rm{disk}}-\rm{SpT}$ scaling 
law to the corresponding relations established for higher mass stars.




\section{Summary}
\label{sec:summary}
We re-processed the {\it Herschel}/PACS data and measured the 70 and $160\,\mu{\rm{m}}$ flux density for a sample of 
55 BDs and very low mass stars with SpT ranging from M5.5 to L0. Supplemented with previous observations 
at shorter and (sub-)millimeter wavelengths, we constructed their broadband SEDs with extended wavelength coverage, 
providing a valuable opportunity to characterize the disk properties in the substellar mass regime. 
We divided our sample into two groups, i.e. an ETBD and LTBD group that consists of targets with SpT 
earlier or later than M8. We found significant difference in the cumulative distributions of the flux densities between these
two groups. In particular, the ETBD group features a median $70\,\mu{\rm{m}}$ flux density at 100\,pc that is at least 
three times higher than that of the other group. We studied the dependence of disk properties on the SpT
(as a proxy for the mass) of the central object down to the low-mass BD regime and even entering the planetary 
regime. This knowledge is important for us to understand the formation mechanism of BDs and to improve 
the planet formation models. Unlike previous studies, our analysis is homogeneous all the way from {\it Herschel}/PACS data 
reduction and flux density measurement to SED modeling within the substellar regime.

For the 46 objects that show IR excess above the photospheric emission, we performed detailed SED analysis 
using self-consistent radiative transfer models. We introduced as few free parameters as possible in the 
fitting procedure in order to reduce the model degeneracy. The statistics based on the entire sample shows 
that the disk flaring of BDs and very low mass stars is indeed smaller than that of their higher mass counterparts 
like T Tauri disks, the disk mass is orders of magnitude lower than the typical value found in T Tauri stars, and 
the scale height is independent on the mass of the central object. Moreover, we systematically compared the modeling
results from Bayesian analysis between the ETBD and LTBD groups and found similar trends of flaring index  
as a function of SpT. The disk scale heights are comparable in both the high-mass and very low-mass
BDs. Both groups feature a similar median disk mass and no clear trend is visible in the distribution, probably 
due to the uncertainty in translating the PACS photometry into disk mass, the detection bias, and the difference 
in evolutionary stage among the targets. Future deeper far-IR surveys and ALMA observations are required to improve 
constraints on the underlying morphology of the relationship between the stellar and disk properties in the low-end 
stellar mass regime.

\begin{acknowledgements}
We acknowlegde helpful discussions about Herschel/PACS observations with Ulrich Klaas and Hendrik Linz.
Y.L. acknowledges the support by the Natural Science Foundation of Jiangsu Province (Grant No. BK20141046).
This work is supported by the Strategic Priority Research Program ``The Emergence of Cosmological Structures'' 
of the Chinese Academy of Sciences, Grant No. XDB09000000. A.B. acknowledges financial support from the 
Proyecto Fondecyt de Iniciaci\'on 11140572. M.N. is funded by the Deutsches Zentrum f\"ur Luft- und 
Raumfahrt (DLR). This publication makes use of data products from 
the Wide-field Infrared Survey Explorer, which is a joint project of the University of California, Los Angeles, 
and the Jet Propulsion Laboratory/California Institute of Technology, funded by the National Aeronautics 
and Space Administration. PACS has been developed by a consortium of institutes led by MPE (Germany) and 
including UVIE (Austria); KU Leuven, CSL, IMEC (Belgium); CEA, LAM (France); MPIA (Germany); 
INAF- IFSI/OAA/OAP/OAT, LENS, SISSA (Italy); IAC (Spain).
\end{acknowledgements}

\bibliographystyle{aa}
\bibliography{vlmbdref}

\newpage
\begin{appendix}
\section{The Bayesian probability distributions of selected disk parameters for each modeled object}
\begin{figure}[!h]
\includegraphics[width=0.5\textwidth]{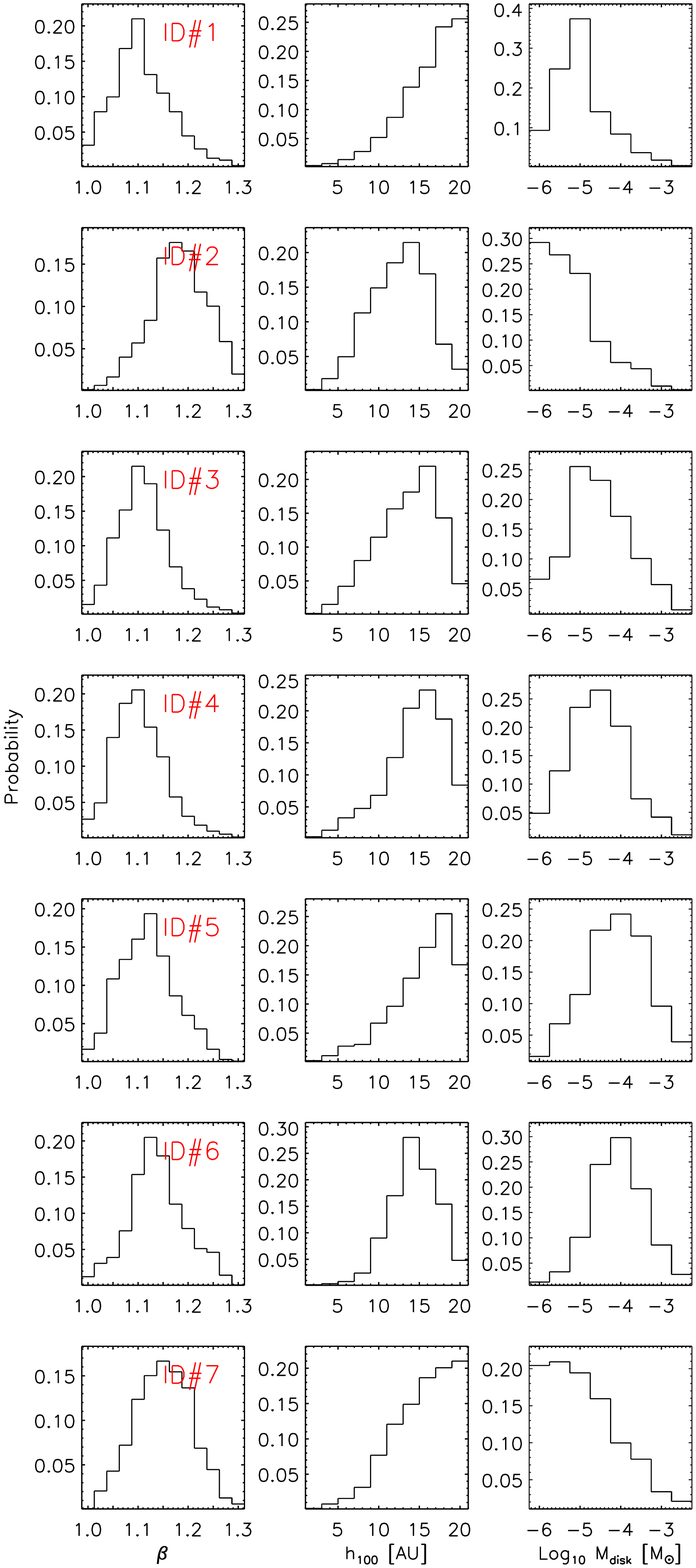}
\caption{Bayesian probability distributions of $\beta$, $H_{100}$ and $M_{\rm disk}$ for each object.}
\end{figure}

\addtocounter{figure}{-1}
\begin{figure}[!t]
\vspace*{2.2cm}
\includegraphics[width=0.5\textwidth]{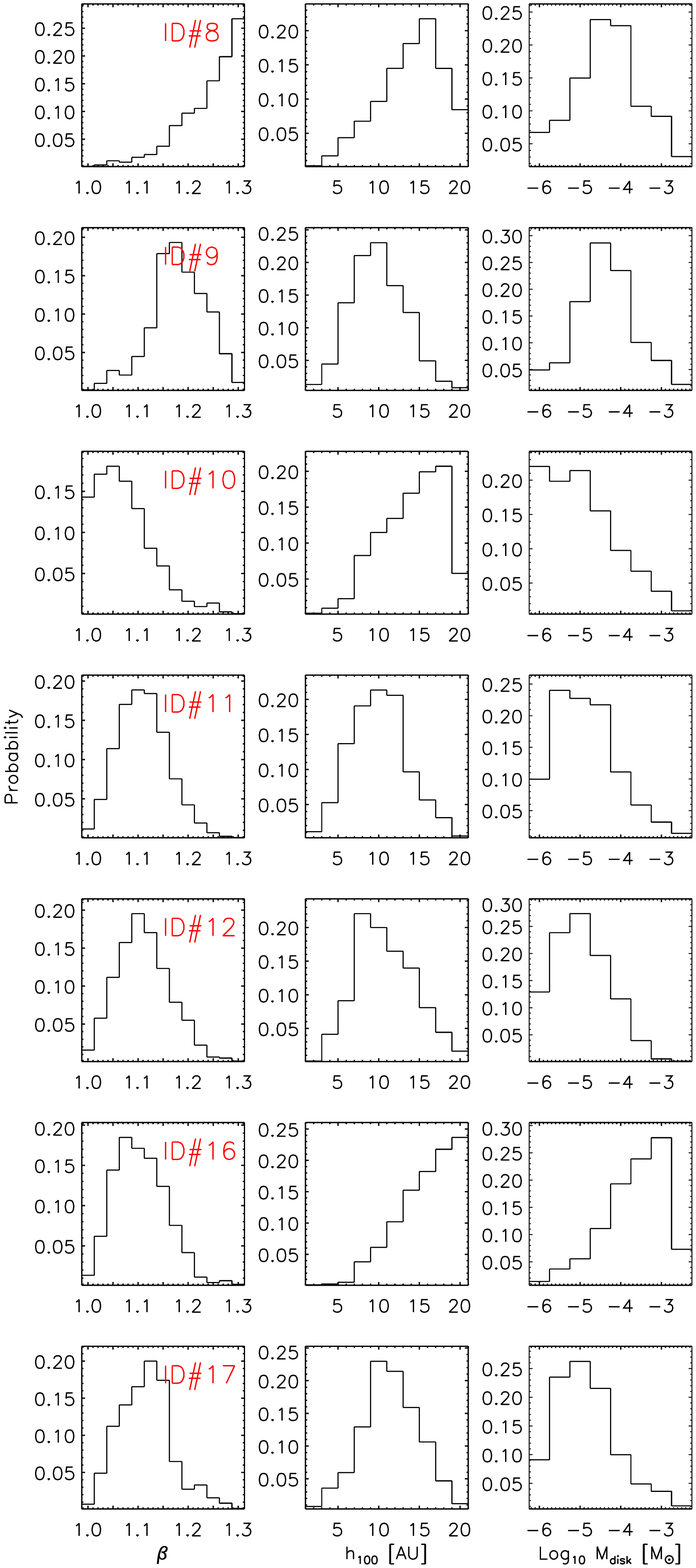}
\caption{Continued.}
\end{figure}

\addtocounter{figure}{-1}
\begin{figure}[!t]
\includegraphics[width=0.5\textwidth]{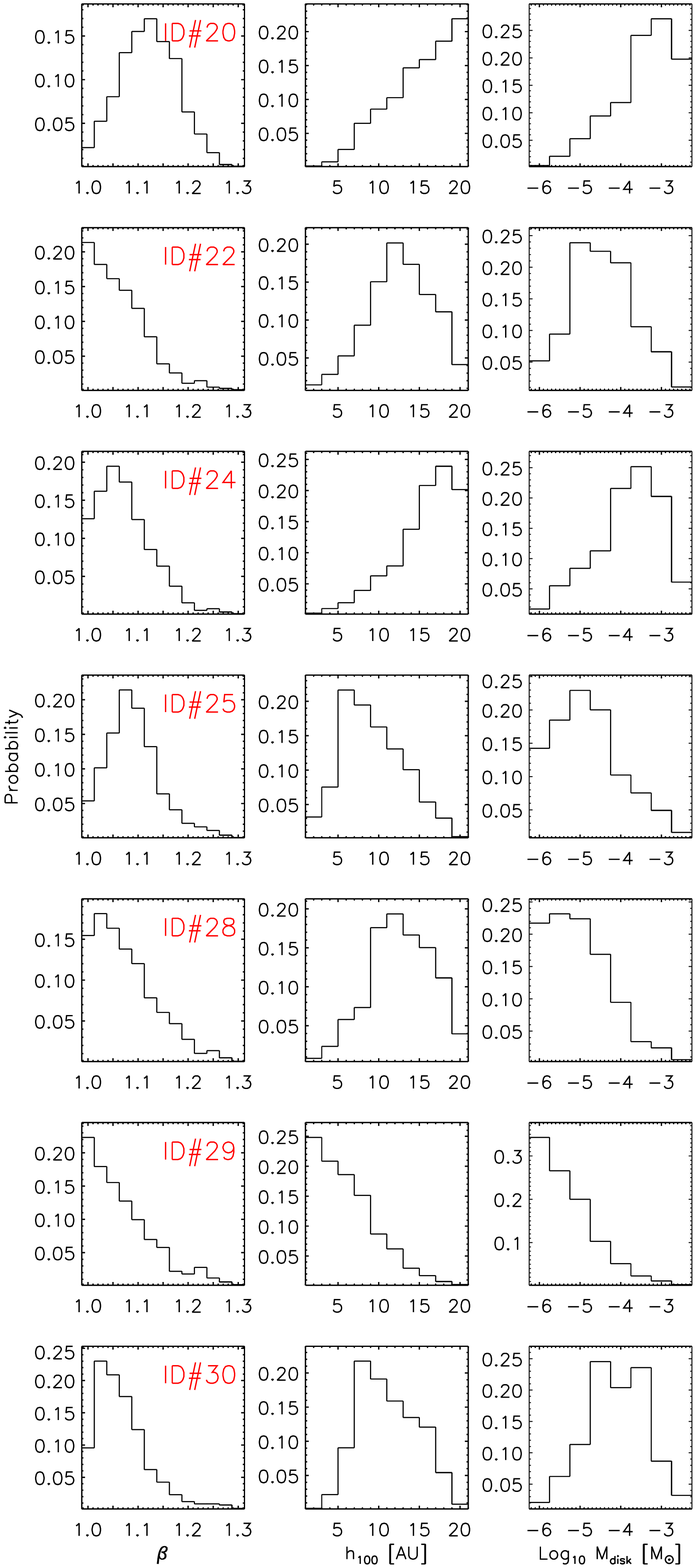}
\caption{Continued.}
\end{figure}

\addtocounter{figure}{-1}
\begin{figure}[!t]
\includegraphics[width=0.5\textwidth]{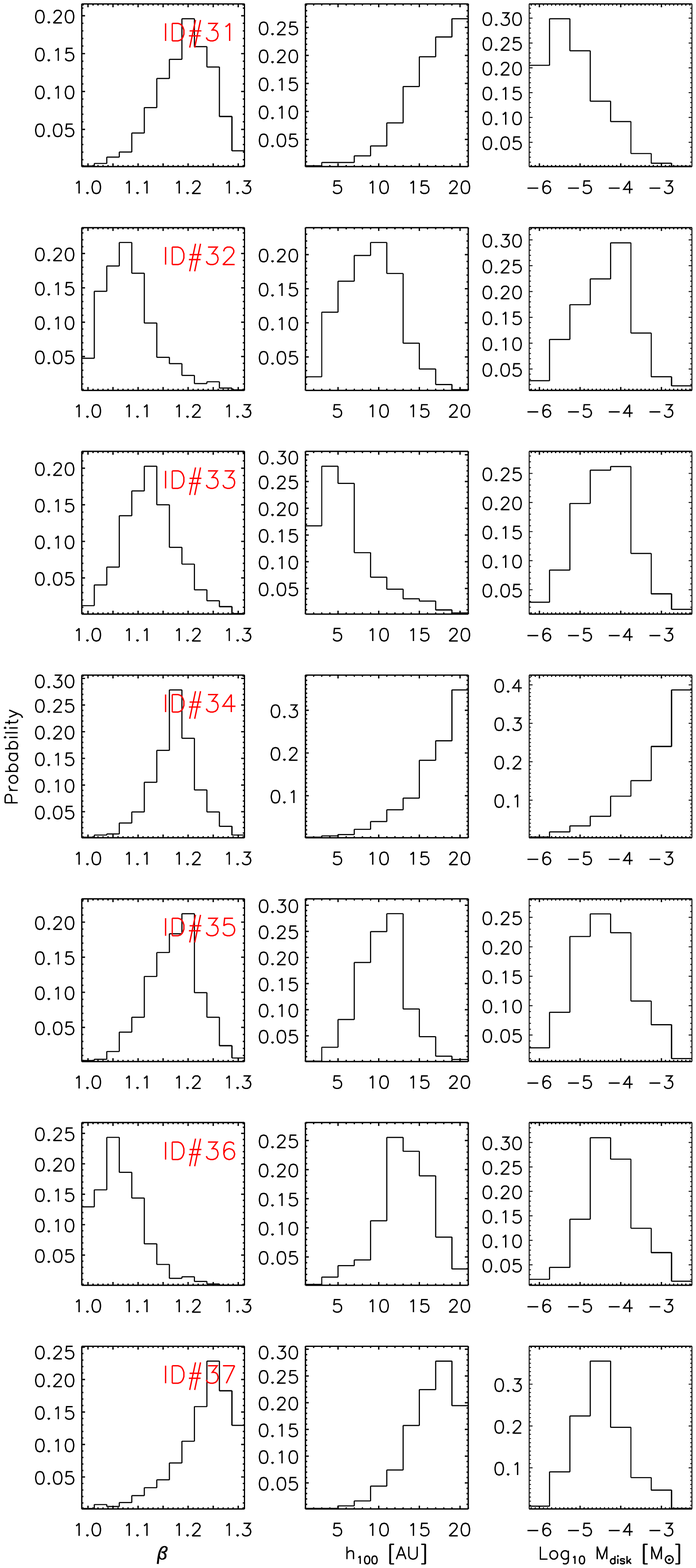}
\caption{Continued.}
\end{figure}

\addtocounter{figure}{-1}
\begin{figure}[!t]
\includegraphics[width=0.5\textwidth]{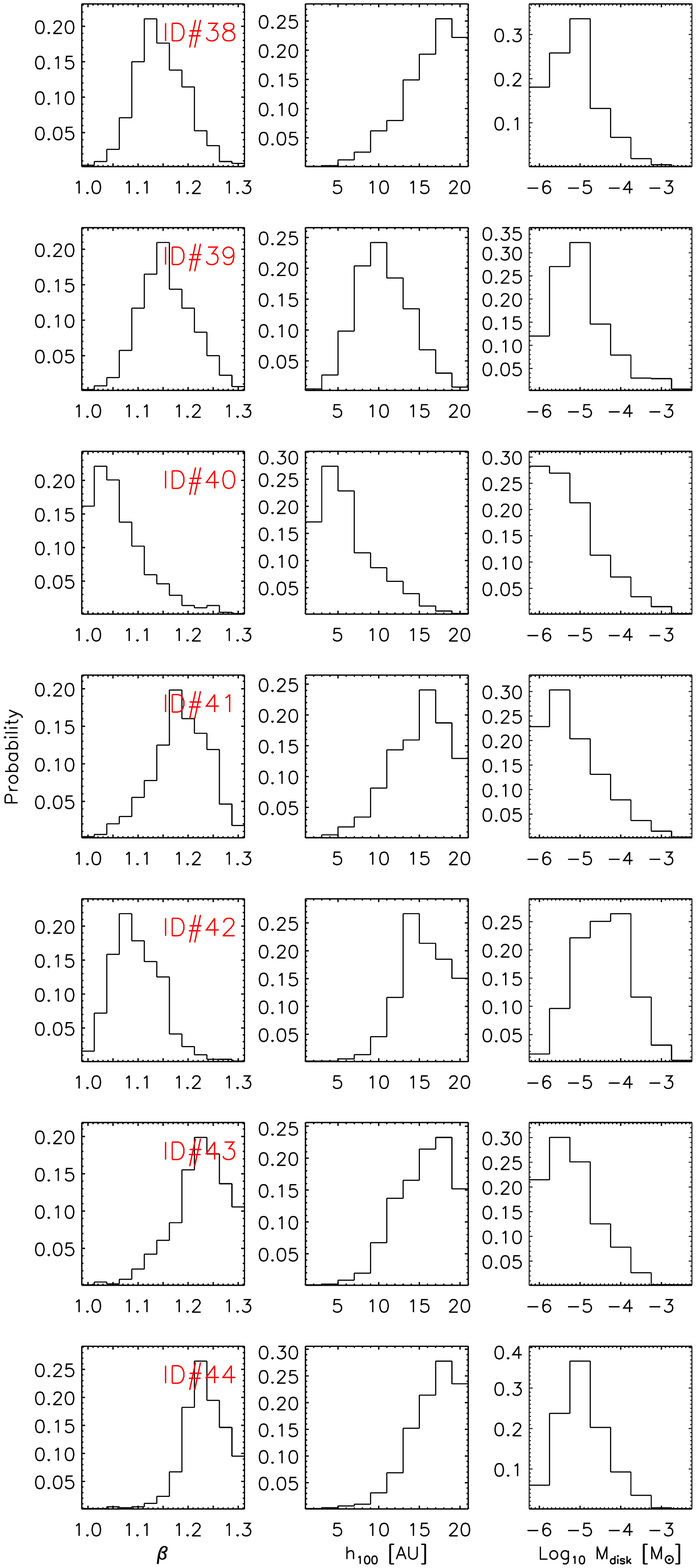}
\caption{Continued.}
\end{figure}

\addtocounter{figure}{-1}
\begin{figure}[!t]
\includegraphics[width=0.5\textwidth]{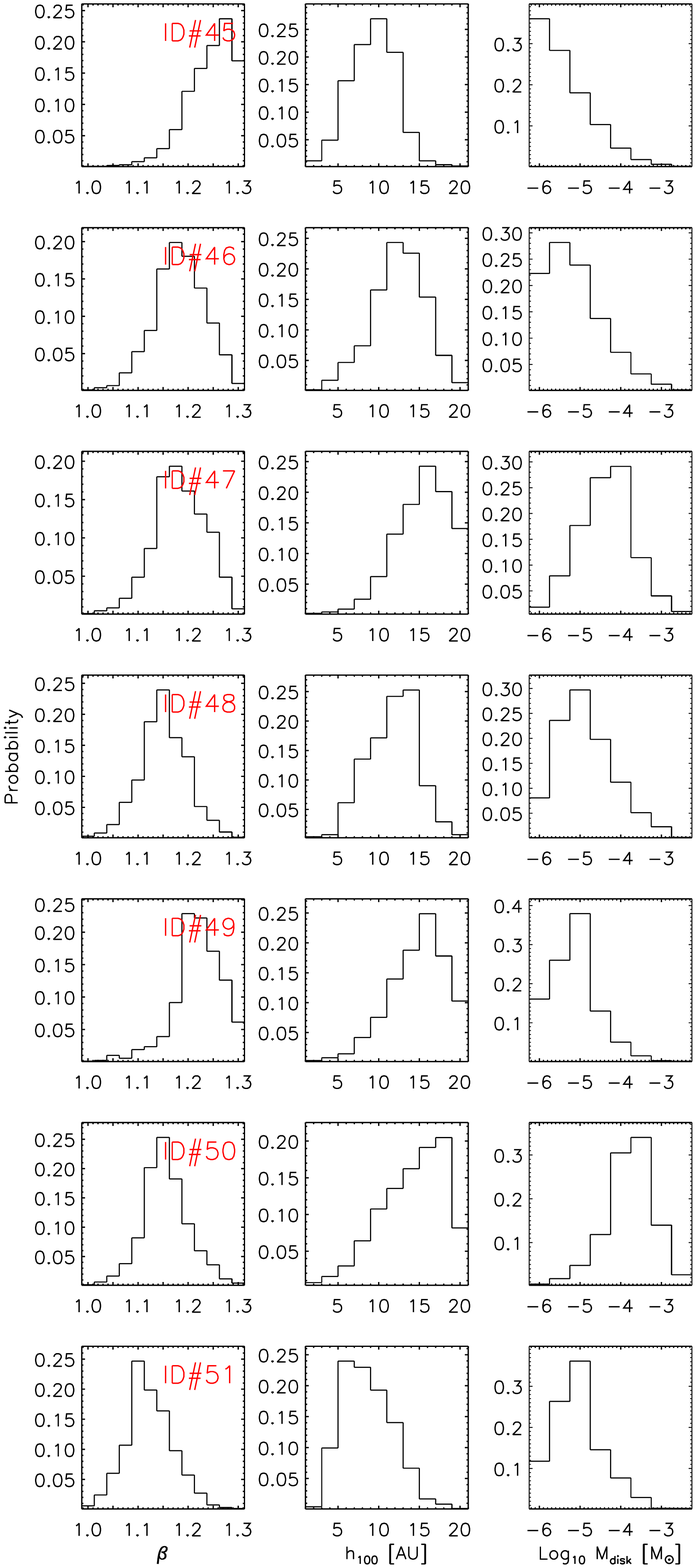}
\caption{Continued.}
\end{figure}

\addtocounter{figure}{-1}
\begin{figure}[!t]
\includegraphics[width=0.5\textwidth]{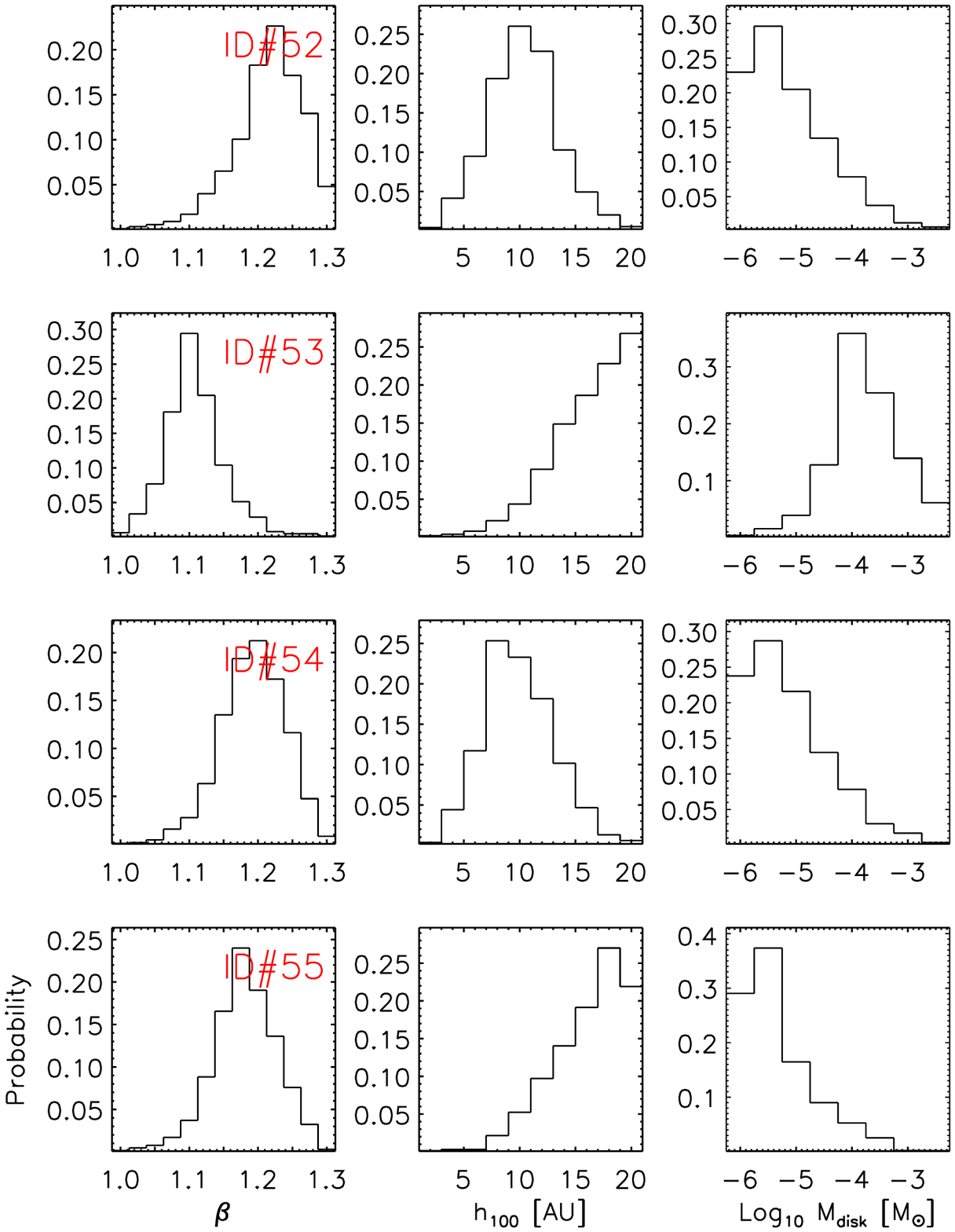}
\caption{Continued.}
\end{figure}

\end{appendix}
\end{document}